\documentclass[journal=jctcce,manuscript=article]{achemso}
\setkeys{acs}{articletitle = true}
\setkeys{acs}{chaptertitle = true}
\setkeys{acs}{abbreviations = false}
\setkeys{acs}{maxauthors=150}
\setkeys{acs}{etalmode=truncate}
\usepackage{amsmath}
\usepackage{amssymb}
\usepackage[version=4]{mhchem}

\usepackage{graphicx}
\usepackage{multirow}
\usepackage{siunitx}
\usepackage{xcolor}
\usepackage{braket}
\usepackage{bm}
\usepackage{amsmath}

\usepackage[hidelinks]{hyperref}
\hypersetup{colorlinks=false}

\usepackage{numprint}
\usepackage[utf8]{inputenc}
\usepackage[T1]{fontenc}
\usepackage{mathptmx}
\usepackage[english]{babel}
\usepackage{subfigure}
\usepackage{comment}
\usepackage{siunitx}
\usepackage{datetime}
\newdate{mydate}{21}{12}{2023}

\title{Which options exist for NISQ-friendly linear response formulations?} 

\author{Karl Michael Ziems}
\email{kmizi@kemi.dtu.dk}
\affiliation{Department of Chemistry, Technical University of Denmark, Kemitorvet Building 207, DK-2800 Kongens Lyngby, Denmark.}

\author{Erik Rosendahl Kjellgren}
\email{kjellgren@sdu.dk}
\affiliation{Department of Physics, Chemistry and Pharmacy,
University of Southern Denmark, Campusvej 55, 5230 Odense, Denmark.}

\author{Peter Reinholdt}
\affiliation{Department of Physics, Chemistry and Pharmacy,
University of Southern Denmark, Campusvej 55, 5230 Odense, Denmark.}

\author{Phillip W. K. Jensen}
\affiliation{Department of Chemistry, University of Copenhagen, DK-2100 Copenhagen \O.}
\author{Stephan P. A. Sauer}
\affiliation{Department of Chemistry, University of Copenhagen, DK-2100 Copenhagen \O.}
\author{Jacob Kongsted}
\affiliation{Department of Physics, Chemistry and Pharmacy,
 University of Southern Denmark, Campusvej 55, 5230 Odense, Denmark.}
\author{Sonia Coriani}
\affiliation{Department of Chemistry, Technical University of Denmark, Kemitorvet Building 207, DK-2800 Kongens Lyngby, Denmark.}
\email{soco@kemi.dtu.dk}

\date{\displaydate{mydate}}

\begin{document}
%TO REMEMBER WE NEED THIS
%\begin{tocentry}
%\end{tocentry}

\begin{abstract}
Linear response (LR) theory is a powerful tool in classic quantum chemistry crucial to understanding photo-induced processes in chemistry and biology. However, performing simulations for large systems and in the case of strong electron correlation remains challenging. Quantum computers are poised to facilitate the simulation of such systems, and recently, a quantum linear response formulation (qLR) was introduced. To apply qLR to near-term quantum computers beyond a minimal basis set, we here introduce a resource-efficient qLR theory using a truncated active-space version of the multi-configurational self-consistent field LR ansatz. Therein, we investigate eight different near-term qLR formalisms that utilize novel operator transformations that allow the qLR equations to be performed on near-term hardware. Simulating excited state potential energy curves and absorption spectra for various test cases, we identify two promising candidates dubbed 
``proj LRSD'' and ``all-proj LRSD''.

\end{abstract}

\maketitle

\section{\label{sec:intro}Introduction}

Theoretical modeling and computational simulation of the response of matter to light provide a fundamental understanding of the microscopic mechanisms behind all types of spectroscopy as well as of light-energy conversion processes with a vital impact on chemistry, life sciences, and material design.
Over the past five decades, tremendous progress has been made in the development of the theoretical framework known as molecular response theory~\cite{olsen1985linear,Christiansen1998,helgaker2012recent,Pawlowski2015} to enable the computation of molecular response properties and spectra on ``classical'' hardware. 
The framework is nowadays implemented in several general-purpose quantum chemistry packages (see, e.g., Refs.~\citenum{Aidas2013,DALTON2022,Qchem541,Turbomole2023,Psi4,eTprog,veloxchem}) for a variety of wave-function parametrizations as well as within time-dependent density-functional theory, and it encompasses both linear and non-linear response properties and spectra.~\cite{Christiansen1998,Salek2002,Norman2011perspective,helgaker2012recent} 

Essential ingredients in all these formulations are the solution of a (generalized) eigenvalue problem to determine excitation energies and excitation vectors and the solution of linear response equations to determine the wave-function/density-function response parameters.~\cite{olsen1985linear,
helgaker2012recent}
Once these fundamental building blocks are in place, it becomes relatively straightforward to calculate different types of response properties and spectra for the chosen parametrization of the ground state. 

However, challenges remain regarding how to design the most accurate approaches, which also have the steepest scaling, into computationally amenable ones (with reasonable execution times)  for large molecular systems with challenging (strongly correlated) electronic structures.

In the last decade, quantum computing technology has been capturing the imagination of the scientific world.~\cite{Zelevinsky2023} Quantum processing units (QPUs) exploit the quantum-mechanical phenomena of superposition and entanglement to perform a calculation. 
Still limited hardware-wise, it is believed that they have the potential, in due time, to revolutionize high-performance computing; especially quantum chemistry has long been pointed out as an obvious area of application of quantum computing.\cite{Aspuru-Guzik2005,Cao2019,Bauer2020}
% However, before quantum computing can be truly utilized, new theories and tailor-made algorithms are required. 

While fault-tolerant quantum computers are still in the making, the current paradigm to solve chemistry problems on near-term quantum hardware (commonly referred to as noisy intermediate-scale quantum, NISQ~\cite{chen_complexity_2023}) is via a hybrid approach involving quantum and classical computers working in tandem.~\cite{McClean2016}   
The NISQ device is used to perform measurements of specific expectation values or density matrix elements while the rest of the calculation is offloaded to the classical processors.
%Quantum computing in quantum chemistry is in many ways still in its infancy and hardly extends beyond proof-of-principle.
For hybrid computations on NISQ devices to go beyond proof of principle and possibly become competitive with traditional electronic structure theory for classical computers, it is crucial to develop ``computationally 
frugal'' algorithms, i.e., algorithms utilizing a small number of qubits and characterized by a modest circuit depth. 

After a few years where the focus of NISQ computations in quantum chemistry has been on ground-state preparation, the attention is nowadays progressively shifting towards other properties than ground-state energies, in particular, towards linear response (LR) properties, such as excitation energies, transition strengths, and general linear response functions.~\cite{Ollitrault2020,Asthana2023,Kumar2023,Kim2023,Castellanos2023,nakagawa_analytical_2023,Jensen_qEOM_2023} 
% To this end, it becomes paramount to develop new
% theories and 
% tailor-made algorithms that can harness the potential of NISQ

In this work, we aim to formulate a resource-efficient, time-dependent ``quantum'' linear response (qLR) theory with truncated active spaces suitable for computations on near-term quantum computers.
Resorting to active spaces allows us to treat larger basis sets than so far attempted on (or emulated for) noisy quantum computers.
Moreover, truncating the active space to a certain excitation rank in the linear response can be helpful for less converged wavefunctions, as it puts less stress on how (well) parametrized the underlying wave function ansatz is. Both lead to fewer parameters and, in turn, to a smaller Hessian matrix to be measured.

Using the multi-configurational self-consistent field (MCSCF) linear-response ansatz~\cite{olsen1985linear,helgaker2012recent} as a starting point, we define eight qLR ansatzes by different choices of the orbital rotation and (active-space) excitation operators in the response equations.
These alternative operator choices are defined through specific transformations (one trivial and three non-trivial) of the naive operators of traditional MCSCF response theory.
The general suitability for near-term application of each qLR ansatz will be evaluated considering the number and circuit depth of the unique terms that should be measured on the quantum device. 
The impact of the alternative transformations of the active-space excitation and the orbital rotation operators, together with the truncation of the active excitation space, is investigated by computing excitation energies and oscillator strengths for selected cases and comparing them with the results of traditional complete-active space self-consistent field (CASSCF) and full configuration interaction (FCI) calculations. 
%\sonia{Note that the }

%becomes relevant.
%\vspace{\baselineskip}

% Why our qLR approach:
% \begin{itemize}
%     \item active spaces allow much larger basis sets than currently done on QC (we use up to cc-pVDZ)
%     \item truncation of active space helpful for less converged wf 
%     \item truncation of active space leads to smaller Hessian
%     \item truncation of active space needs fewer parameters
%     \item But influence of transformation of active space excitation and orbital rotation becomes relevant
% \end{itemize}

%IF NEEDED WE CAN HAVE A SUMMARY
%The paper is organized as follows:

\section{\label{sec:theo}Theory}
As anticipated in the previous section, this work formulates a time-dependent LR theory with active spaces suitable for near-term quantum computers. 
To this end, we first introduce the orbital-optimized unitary coupled cluster (oo-UCC) ansatz in an active space, followed by our qLR theory within a truncated active-space multi-configurational self-consistent field framework. 
Hereby, we focus on various operator formalisms for LR. We analyze these regarding their feasibility on near-term quantum computers in the last section.

Throughout the paper, the index convention is that $p,q,r,s$ are general indices, $a,b,c,d$ are virtual indices, $i,j,k,l$ are inactive indices, and $v,w,x,y$ are active indices.
The notation $v_a$ is an active index that is virtual in the Hartree-Fock reference, and $v_i$ is an active index that is inactive in the Hartree-Fock reference.

\subsection{Orbital-optimized unitary coupled cluster, oo-UCC
\label{sec:theo:ooUCC}} 
The orbital-optimized unitary coupled cluster (oo-UCC) method is a recently developed, chemistry-inspired ansatz for the wave function on a quantum computer.~\cite{Mizukami2020,Sokolov2020,Fitzpatrick2022} Its classic counterpart is the wave function obtained via the multi-configurational self-consistent field (MCSCF) method.~\cite{Helgaker2013book}
Just like MCSCF, oo-UCC splits the full space into smaller subspaces. In this work, the full space is split into three parts: inactive orbitals, active orbitals, and virtual orbitals.
The complete active space (CAS) method uses the same partitioning.~\cite{Siegbahn1980,Roos1980,Siegbahn1981}
While CAS implies a complete CI expansion is performed within the active space, we here truncate the excitation expansion in the active space; hence, we will refer to the active space partition as ``CAS-like''.
Applying a CAS-like active space approximation of the wave function leads to the following decomposition:
\begin{equation}
    \left|0\left(\boldsymbol{\theta}\right)\right> = \left|I\right>\otimes \left|A\left(\boldsymbol{\theta}\right)\right> \otimes \left|V\right>~,
\end{equation}
with $\boldsymbol{\theta}$ being the parameterization of the active space, $\left|I\right>$ denoting the inactive part, $\left|A\left(\boldsymbol{\theta}\right)\right>$ the active part, and $\left|V\right>$ the virtual part.
The active part is the only subspace in which a unitary transformation from an ansatz is applied, $\left|A\left(\boldsymbol{\theta}\right)\right>=\boldsymbol{U}\left(\boldsymbol{\theta}\right)\left|A\right>$.
Similarly to the wave function, any operator can be decomposed as 
\begin{equation}
    \hat{O} = \hat{O}_I\otimes \hat{O}_A\otimes \hat{O}_V~.
\end{equation}
This yields expectation values of the form:
\begin{equation}
    \left<0\left(\boldsymbol{\theta}\right)\left|\hat{O}\right|0\left(\boldsymbol{\theta}\right)\right> = \left<I\left|\hat{O}_{I}\right|I\right>\otimes \left<A\left(\boldsymbol{\theta}\right)\left|\hat{O}_{A}\right|A\left(\boldsymbol{\theta}\right)\right> \otimes\left<V\left|\hat{O}_{V}\right|V\right>~. \label{eq:op_splitting}
\end{equation}
The inactive part $\left<I\left|\hat{O}_{I}\right|I\right>$ and the virtual part $\left<V\left|\hat{O}_{V}\right|V\right>$ are trivial to calculate and any computationally demanding part is directly related to the size of the active space. Therefore, in quantum simulations with active spaces, the inactive and virtual parts are simulated classically, while the active part is translated to quantum hardware.

The ansatz used to parameterize the active space and the orbitals for the oo-UCC wave function is
\begin{equation}
    \left|\text{oo-UCC}\right> = \exp\left\{-\hat{\kappa}\left(\boldsymbol{\kappa}\right)\right\}\exp\left\{-\hat{t}\left(\boldsymbol{\theta}\right)\right\}\left|\text{CSF}\right>
\end{equation}
with the orbital rotation parameter being
\begin{equation}
    \hat{\kappa}\left(\boldsymbol{\kappa}\right) = \sum_{pq}\kappa_{pq}\hat{E}^-_{pq}~.
\end{equation}
Here, $\hat{E}^-_{pq} = \hat{E}_{pq} - \hat{E}_{qp}$, where $\hat{E}_{pq} = \hat{a}_{p,\alpha}^\dagger\hat{a}_{q,\alpha} + \hat{a}_{p,\beta}^\dagger\hat{a}_{q,\beta}$ is the singlet single-excitation operator, and $\left|\text{CSF}\right>$ being a single configuration state function reference.

The only non-redundant orbital rotations when combined with UCC are rotations of type \textit{inactive to active}, \textit{inactive to virtual}, and \textit{active to virtual}, $pq\in\left\{ai,av,vi\right\}$.
The unitary coupled cluster operator, $\hat{t}(\boldsymbol{\theta})$, acts within the active space and reads
\begin{equation}
    \hat{t}(\boldsymbol{\theta}) = \hat{T}(\boldsymbol{\theta}) - \hat{T}^\dagger(\boldsymbol{\theta})
\end{equation}
with the cluster operator, $\hat{T}(\boldsymbol{\theta})$, defined as
\begin{equation}
    \hat{T}(\boldsymbol{\theta}) = \sum_k\ ^{(k)}\hat{T}(\boldsymbol{\theta}). \label{eq:cluster_op}
\end{equation}

For the UCC singles and doubles (UCCSD) wave function, $k=\{1,2\}$ in \autoref{eq:cluster_op}, and we can write the singles and doubles cluster operators as
\begin{equation}
    \ ^{(1)}\hat{T}(\boldsymbol{\theta}) = \sum_{v_I v_A} \theta_{v_I}^{v_A}\hat{a}^\dagger_{v_A} \hat{a}_{v_I}\label{eq:ucc_single}
\end{equation}
and
\begin{equation}
    \ ^{(2)}\hat{T}(\boldsymbol{\theta}) = \sum_{\substack{v_I<v_J\\v_A<v_B}} \theta_{v_Iv_J}^{v_Av_B}\hat{a}^\dagger_{v_A}\hat{a}_{v_I} \hat{a}^\dagger_{v_B} \hat{a}_{v_J}\label{eq:ucc_double}.
\end{equation}

Instead of letting the orbital rotation parametrization work on the state vector, it is used to transform the integrals in the Hamiltonian,
\begin{equation}
    \hat{H}\left(\boldsymbol{\kappa}\right) = \sum_{pq}h_{pq}\left(\boldsymbol{\kappa}\right)\hat{E}_{pq} + \frac{1}{2}\sum_{pqrs}g_{pqrs}\left(\boldsymbol{\kappa}\right)\hat{e}_{pqrs}~.
\end{equation}
Here, $\hat{e}_{pqrs} = \hat{E}_{pq}\hat{E}_{rs} - \delta_{qr}\hat{E}_{ps}$ is the two-electron excitation operator.
This transforms the integrals to
\begin{align}
h_{pq}\left(\boldsymbol{\kappa}\right) &= \sum_{p'q'} \left[\exp\left(\boldsymbol{\kappa}\right)\right]_{q'q}h_{p'q'}\left[\exp\left(-\boldsymbol{\kappa}\right)\right]_{p'p}\\
g_{pqrs}\left(\boldsymbol{\kappa}\right) &= \sum_{p'q'r's'}\left[\exp\left(\boldsymbol{\kappa}\right)\right]_{s's}\left[\exp\left(\boldsymbol{\kappa}\right)\right]_{q'q}g_{p'q'r's'}\left[\exp\left(-\boldsymbol{\kappa}\right)\right]_{p'p}\left[\exp\left(-\boldsymbol{\kappa}\right)\right]_{r'r}.
\end{align}

The energy of the ground state can now be found by variational minimization in the parameters $\boldsymbol{\theta}$ and $\boldsymbol{\kappa}$,
\begin{equation}
    E_\text{gs} = \min_{\boldsymbol{\theta}, \boldsymbol{\kappa}}\left<\text{UCC}\left(\boldsymbol{\theta}\right)\left|\hat{H}\left(\boldsymbol{\kappa}\right)\right|\text{UCC}\left(\boldsymbol{\theta}\right)\right>.
\end{equation}
Here, $\left|\text{UCC}\right> = \exp\left(-\hat{t}\left(\boldsymbol{\theta}\right)\right)\left|\text{CSF}\right>$. Performing this minimization process on quantum architecture is known as the orbital-optimized variational quantum eigensolver (oo-VQE) algorithm.\cite{Mizukami2020,Sokolov2020,Fitzpatrick2022} Its major advantage is splitting the problem into the active space simulations on the quantum computer and performing all operator applications outside the active space, i.e., \ the orbital rotations, classically.

\subsection{\label{sec:theo:LR}
Quantum linear response theory with active spaces}

In the following, $\left|0\right>$ refers to any reference wave function. For the numerical examples discussed in this work, the reference will be of the type $\left|\text{UCC}\right>$ as introduced above.
Note, however, that any other variational ansatz, such as ADAPT~\cite{Grimsley2019} or qubit-ADAPT,~\cite{Tang2021}, can also be used without changing the linear response equations.

\subsubsection{Basics of classic Linear Response}

In classic quantum chemistry, linear response with active spaces is performed using the MCSCF framework. The time-dependent MCSCF wave function, $\ket{\tilde 0}$, can be obtained from the unperturbed MCSCF reference wave function, $\ket{0}$, using the parametrization
\begin{equation}
    \left|{\tilde 0}\right> = \left.\left.\exp\left\{i\hat\kappa(t)\right\}\exp\left\{i\hat S(t)\right\}\right|0\right>.
    \label{Otilde}
\end{equation}
In \autoref{Otilde}, $\hat \kappa(t)$ and $\hat S(t)$ are the time-dependent
Hermitian orbital and active-space rotation operators, respectively,
\begin{align}
    \hat{\kappa}(t) &= \sum_\mu \left( \kappa_\mu(t) \hat{Q}_\mu + \kappa_\mu^*(t)\hat{Q}_\mu^\dagger \right) \label{eq:kappa_t} \\
    \hat{S}(t) &= \sum_n \left( S_n(t) \hat{R}_n + S_n^*(t)\hat{R}_n^\dagger \right) \label{eq:S_t}
\end{align}
containing the generic orbital rotation operator, $\hat{Q}_\mu$, and the generic active-space excitation operator, $\hat{R}_n$. 
The exact definition of $\hat{Q}_\mu$ and $\hat{R}_n$ will be addressed shortly in \autoref{sec:transform}.
Following established derivations, either based on the time-dependent expectation value of a time-independent operator,~\cite{olsen1985linear} or on the time-averaged MCSCF quasi-energy,~\cite{helgaker2012recent} and expansion of \autoref{eq:kappa_t} and \ref{eq:S_t} in first order, the  MCSCF linear response equation is obtained.
In matrix notation for a given perturbation $\hat{B}$ at frequency $\omega_B$, this takes the form
\begin{align}
 \left( \textbf{E}^{[2]} - \omega_B \textbf{S}^{[2]} \right) \boldsymbol{\beta}_B= i \textbf{V}^{[1]}_B \label{eq:LR}
\end{align}
where $\textbf{E}^{[2]}$ is the electronic Hessian, $\textbf{S}^{[2]}$ is the Hermitian metric matrix, $\boldsymbol{\beta}_B$ is the linear response (column) vector, and $\textbf{V}^{[1]}_B$
is the perturbed gradient
(or property gradient) column vector. 
The Hessian and metric matrices are defined as follows:
\begin{align}
 \textbf{E}^{[2]} &= \begin{pmatrix}
    \boldsymbol{A} & \boldsymbol{B}          \\[0.3em]
      \boldsymbol{B}^* & \boldsymbol{A}^*           
     \end{pmatrix}, \quad 
     \textbf{S}^{[2]} = \begin{pmatrix}
    \boldsymbol{\Sigma} & \boldsymbol{\Delta}          \\[0.3em]
     -\boldsymbol{\Delta} ^* &  -\boldsymbol{\Sigma}^*           
     \end{pmatrix}, \quad
\end{align}
with the submatrices
\begin{align}
    \boldsymbol{A} &= \begin{pmatrix}
\left<0\left|\left[\hat{Q}_\mu^\dagger,\left[\hat{H},\hat{Q}_{\mu^\prime}\right]\right]\right|0\right>
& \left<0\left|\left[\hat{R}_{n^\prime},\left[\hat{H},\hat{Q}^\dagger_\mu\right]\right]\right|0\right> \\
\left<0\left|\left[\hat{R}_{n}^\dagger,\left[\hat{H},\hat{Q}_{\mu^\prime}\right]\right]\right|0\right>
& \left<0\left|\left[\hat{R}_{n}^\dagger,\left[\hat{H},\hat{R}_{n^\prime}\right]\right]\right|0\right>
\end{pmatrix},\\
    \boldsymbol{B} &= \begin{pmatrix}
\left<0\left|\left[\hat{Q}_\mu^\dagger,\left[\hat{H},\hat{Q}_{\mu^\prime}^\dagger\right]\right]\right|0\right>
& \left<0\left|\left[\hat{R}^\dagger_{n^\prime},\left[\hat{H},\hat{Q}^\dagger_\mu\right]\right]\right|0\right> \\
\left<0\left|\left[\hat{R}_{n}^\dagger,\left[\hat{H},\hat{Q}_{\mu^\prime}^\dagger\right]\right]\right|0\right>
& \left<0\left|\left[\hat{R}_{n}^\dagger,\left[\hat{H},\hat{R}_{n^\prime}^\dagger\right]\right]\right|0\right>
\end{pmatrix},\\
    \boldsymbol{\Sigma} &= \begin{pmatrix}
\left<0\left|\left[\hat{Q}_\mu^\dagger,\hat{Q}_{\mu^\prime}\right]\right|0\right>
& \left<0\left|\left[\hat{Q}_{\mu^\prime}^\dagger,\hat{R}_{n}\right]\right|0\right> \\
\left<0\left|\left[\hat{R}_{n}^\dagger,\hat{Q}_{\mu^\prime}\right]\right|0\right>
& \left<0\left|\left[\hat{R}_{n}^\dagger,\hat{R}_{n^\prime}\right]\right|0\right>
\end{pmatrix},\\
    \boldsymbol{\Delta} &= \begin{pmatrix}
\left<0\left|\left[\hat{Q}_\mu^\dagger,\hat{Q}_{\mu^\prime}^\dagger\right]\right|0\right>
& \left<0\left|\left[\hat{Q}_{\mu^\prime}^\dagger,\hat{R}_{n}^\dagger\right]\right|0\right> \\
\left<0\left|\left[\hat{R}_{n}^\dagger,\hat{Q}_{\mu^\prime}^\dagger\right]\right|0\right>
& \left<0\left|\left[\hat{R}_{n}^\dagger,\hat{R}_{n^\prime}^\dagger\right]\right|0\right>
\end{pmatrix}.
\end{align}
Moreover, some important matrix properties to note are
\begin{align}
\boldsymbol{A} = \boldsymbol{A}^\dagger, \quad
\boldsymbol{B} = \boldsymbol{B}^\mathrm{T}, \quad
\boldsymbol{\Sigma} = \boldsymbol{\Sigma}^\dagger \quad \text{and} \quad 
\boldsymbol{\Delta} = - \boldsymbol{\Delta}^\mathrm{T}.
\end{align}
The property gradient column vector on the right-hand side of the linear response equation has the form
\begin{align}
     \textbf{V}^{[1]}_B = \begin{pmatrix}
     \braket{0|[\hat{Q},\hat{B}]|0}         \\[0.3em]
     \braket{0|[\hat{R},\hat{B}]|0}             \\[0.3em]
     \braket{0|[\hat{Q}^\dagger,\hat{B}]|0}          \\[0.3em]
     \braket{0|[\hat{R}^\dagger,\hat{B}]|0}             \\[0.3em]
     \end{pmatrix}.
\end{align}
The elements of the linear response vector, $\boldsymbol{\beta}_B$, can be grouped in terms of its excitation and de-excitation part
\begin{align}
    \boldsymbol{\beta}_B = \begin{pmatrix}
    \boldsymbol{Z}_B          \\[0.3em]
      \boldsymbol{Y}_B^*            
     \end{pmatrix} = \begin{pmatrix}
    \boldsymbol{\kappa}_B          \\[0.3em]
      \boldsymbol{S}_B             \\[0.3em]
      \boldsymbol{\kappa}_{-B}^*          \\[0.3em]
      \boldsymbol{S}_{-B}^*             \\[0.3em]
     \end{pmatrix}
\end{align}
and contain the first-order orbital rotation and active-space excitation parameters, $\boldsymbol{\kappa}_B$ and $\boldsymbol{S}_B$, respectively, that are based on the first-order expansion of \autoref{eq:kappa_t} and \autoref{eq:S_t}. 
%via the second-order MCSCF quasi-energy and expansion of \autoref{eq:kappa_t} and \ref{eq:S_t} in first order. 
% \\
% \sonia{[You mention the the MCSCF linear response equation, which has now been removed. Having removed it, the reader does not know what the  LR vector is, and the following sentence about the poles is 'hanging'. ]}\kmz{I've been told in the last meeting to remove it as we don't use it for the results. I can add it again :D Where do I mention the LR vector later on? I think I only ever refer to the excitation vector. I also never define the quasi-energy and just mention it as background to the derivation. I dont think all has to be pre-defined. The "problem" with adding the linear response equation again, is that we have to define all its components as well 
% without then ever using these definitions.}

By setting the property gradient vector to zero, which corresponds to finding the poles of the linear response vector, the generalized eigenvalue equation for the electronic excited states of the unperturbed system is obtained
% The poles of the linear response vector define the excited states of the unperturbed system and are obtained via the generalized eigenvalue equation
%
\begin{align}
  \textbf{E}^{[2]}  \boldsymbol{\beta}_k =  \omega_k \textbf{S}^{[2]}\boldsymbol{\beta}_k, \label{eq:LR_nopert}
\end{align}
where  $\omega_k$ is the excitation energy of state $k$, 
%the electronic Hessian, $\textbf{E}^{[2]}$, the Hermitian metric matrix, $\textbf{S}^{[2]}$, 
and  $\boldsymbol{\beta}_k$
is the 
the excitation column vector. 
The elements of the excitation vector, $\boldsymbol{\beta}_k$, can also be grouped in terms of its excitation and de-excitation parts
\begin{align}
    \boldsymbol{\beta}_k = \begin{pmatrix}
    \boldsymbol{Z}_k          \\[0.3em]
      \boldsymbol{Y}_k^*            
     \end{pmatrix} = \begin{pmatrix}
    \boldsymbol{\kappa}_k          \\[0.3em]
      \boldsymbol{S}_k             \\[0.3em]
      \boldsymbol{\kappa}_{-k}^*          \\[0.3em]
      \boldsymbol{S}_{-k}^*             \\[0.3em]
     \end{pmatrix}
\end{align}
that describe excitations with positive eigenvalue, 
$\omega_k >0$, and de-excitations with negative eigenvalue, $\omega_k<0$. 
Their elements are the state-specific first-order orbital rotation and active-space excitation parameters, $\boldsymbol{\kappa}_k$ and $\boldsymbol{S}_k$, respectively, derived from the first-order expansion of \autoref{eq:kappa_t} and \ref{eq:S_t}. 

Having solved the MCSCF linear response equation, \autoref{eq:LR}, for the desired perturbation operator(s), the MCSCF
linear response function can be computed as
% $\braket{\braket{\hat{A};\hat{B}}}_{\omega_B}$, 
%with  is obtained from the operators' respective linear response equations (recall Eq.~\eqref{eq:LR}) and defined as
\begin{align}
\braket{\braket{\hat{A};\hat{B}}}_{\omega_B} &= -i \boldsymbol{\beta}_{-A}^{\dagger}\textbf{V}_B^{[1]} =  -i \boldsymbol{\beta}_{-B}^{\dagger}\textbf{V}_A^{[1]} 
%\\
%    &
= \braket{\braket{\hat{B};\hat{A}}}_{\omega_A},
\end{align}
where 
$\hat{A}$ and $\hat{B}$
are the perturbation operators and 
$\omega_A = - \omega_B$. When $\hat{A}$ and $\hat{B}$ 
correspond to (cartesian components of) the electric dipole operator,
$\hat{\mu}_\gamma$, the electric dipole polarizability is 
obtained,
$\alpha_{\gamma\delta}(\omega) = -\braket{\braket{\hat{\mu}_\gamma,\hat{\mu}_\delta}}_\omega$~.

%%%%%%%%%%%%%%%%%

In this work, we concentrate on the simulation of absorption spectra. This requires, besides the excitation energies from \autoref{eq:LR_nopert}, computing the oscillator strengths, $f_k$, for specific excited states $k$. 
In exact theory, the oscillator strengths in atomic units are given by
\begin{equation}
    f_k = \frac{2}{3}\omega_k \sum_{\gamma=x,y,z} \left| \langle 0|\mu_\gamma |k\rangle\right|^2~.
\end{equation}
%
%\sonia{
The oscillator strengths correspond to the state-specific residues~\cite{olsen1985linear} of the (isotropic) dynamic electric dipole polarizability, $\bar{\alpha} = \frac{1}{3}\textrm{Tr}\{\alpha_{\gamma\gamma}\}$.
%which can be defined in terms of the linear response function of the dipole moment operator, $\hat{\mu}_\gamma$, with $\gamma \in \left\{ x,y,z \right\}$
% \begin{align}
%     \alpha_{\gamma\delta}(\omega) &= -\braket{\braket{\hat{\mu}_\gamma,\hat{\mu}_\delta}}_\omega \label{eq:LRF_alpha}
% \end{align}
This can immediately appreciated by writing the exact sum-over-states expression for the polarizability 
\begin{align}
    \alpha_{\gamma\delta}(\omega) &= \sum_k \frac{\braket{0| \hat{\mu}_\gamma |k}\braket{k | \hat{\mu}_\delta |0}}{\omega-\omega_k} - \sum_k \frac{\braket{0| \hat{\mu}_\delta |k}\braket{k | \hat{\mu}_\gamma |0}}{\omega+\omega_k}~. \label{eq:SOS_alpha}
\end{align}
Defining a normalized excitation operator $\hat{\tilde{O}}_k$ based on the excitation vectors obtained from \autoref{eq:LR_nopert}
\begin{equation}
    \hat{O}_k = \sum_{l \in \mu,n}\left({Z}_{k,l} \hat{X}_l^\dagger + Y_{k,l}\hat{X}_l\right), \label{eq:exc_op}
    \end{equation}
that is, 
    \begin{equation}
    \label{eq:exc_op_normal}
    \hat{\tilde{O}}_k = \frac{\hat{O}_k}{\sqrt{\braket{k|k}}} %\\
    %&=  
    = \frac{\hat{O}_k}{\sqrt{\braket{ 0|[ \hat{O}_k, \hat{O}_k^\dagger ] |0}}}
\end{equation}
with $\hat{X}_l \in \left\{ \hat{Q}_\mu, \hat{R}_n \right\}$, we can rewrite the linear response function expression of 
the polarizability
%Eq.~\eqref{eq:LRF_alpha} 
to resemble the exact sum-over-state expression in \autoref{eq:SOS_alpha}
\begin{align}
    \alpha_{\gamma\delta}(\omega) &= \sum_k \frac{\braket{0| [\hat{\mu}_\gamma,\hat{\tilde{O}}_k] |0}\braket{0 | [\hat{\tilde{O}}_k^\dagger,\hat{\mu}_\delta] |0}}{\omega-\omega_k} - \sum_k \frac{\braket{0| [\hat{\mu}_\delta,\hat{\tilde{O}}_k] |0}\braket{0 | [\hat{\tilde{O}}_k^\dagger,\hat{\mu}_\gamma] |0}}{\omega+\omega_k}. \label{eq:LR_SOS_alpha}
\end{align}
From there, we define the isotropic component 
\begin{align}
    \overline{\alpha}(\omega) = \frac{1}{3} \sum_\gamma \alpha_{\gamma\gamma} = \frac{2}{3} \sum_\gamma \sum_k \frac{\omega_k \left| \braket{0| [\hat{\mu}_\gamma,\hat{\tilde{O}}_k] |0}\right|^2}{\omega^2 - \omega_k^2}
\end{align}
and recognize the state-specific residue as being the oscillator strength
\begin{align}
    f_k &=  \frac{2}{3} \omega_k \sum_\gamma \left| \braket{0| [\hat{\mu}_\gamma,\hat{\tilde{O}}_k] |0}\right|^2. \label{eq:osc_strength}
\end{align}
The dipole moment operator is defined in second quantization as
\begin{align}
    \hat{\mu}_\gamma = - \sum_{pq} \braket{ \phi_p|\vec{r}_\gamma| \phi_q} \hat{E}_{pq}
\end{align}
with $\gamma \in \left\{ x,y,z \right\}$ and $\phi_p$ are the molecular orbitals.
\autoref{eq:osc_strength}
will be the starting point to obtain working equations for $f_k$ within the qLR framework. 

%%%%%%%%%%%%%%%%%%%%
% Within this framework, the oscillator strength, $f_k$, is defined as the state-specific residue of the isotropic dynamic electric dipole polarizibility,
% \begin{align}
%     f_k &=  \frac{2}{3} \sum_\gamma \omega_k \frac{\left| \braket{0| [\hat{\mu}_\gamma,\hat{O}_k] |0}\right|^2}{\sqrt{\braket{ 0|[ \hat{O}_k, \hat{O}_k^\dagger ] |0}}}. \label{eq:osc_strength}
% \end{align}
% The quantity at the denominator, $\sqrt{\braket{ 0|[ \hat{O}_k, \hat{O}_k^\dagger ] |0}}$, is the normalization of the excited state, and the expectation value at the numerator, $\braket{0| [\hat{\mu}_\gamma,\hat{O}_k] |0}$, is the state-specific property gradient with the  excitation operator obtained from \autoref{eq:LR_nopert}
% \begin{equation}
%     \hat{O}_k = \sum_{l \in \mu,n}\left({Z}_{k,l} \hat{X}_l^\dagger + Y_{k,l}\hat{X}_l\right) \label{eq:exc_op}
% \end{equation}
% with $\hat{X}_l \in \left\{ \hat{Q}_\mu, \hat{R}_n \right\}$. The 
% operator (component) $\hat{\mu}_\gamma$ is the electric dipole moment operator, which in second quantization is defined as
% \begin{align}
%     \hat{\mu}_\gamma = - \sum_{pq} \braket{ \phi_p|\vec{r}_\gamma| \phi_q} \hat{E}_{pq}
% \end{align}
% with $\gamma \in \left\{ x,y,z \right\}$ and $\phi_p$ being the molecular 
% orbitals.

\subsubsection{\label{sec:transform}Transformations of operators for qLR}

So far, we have introduced the parametrization of the reference wave function, $\ket{0}$, in terms of the generic orbital rotation operator, $\hat{Q}_\mu$, and the generic active-space excitation operator, $\hat{R}_n$ (recall \autoref{eq:kappa_t} and \ref{eq:S_t}). 
In classic MCSCF theory, these operators are the state-transfer active-space excitation operators with naive orbital rotations (\textit{vide infra}) and span a complete CI expansion in the active space.\cite{olsen1985linear,Jorgensen1988}
However, for our quantum counterpart, qLR, we shall explore novel operator choices in a truncated excitation space to reduce the parameter space. We aim to find methods that can be easily implemented on a near-term quantum device (more in \autoref{sec:qLR_onhardware}). To this end, we introduce \textit{naive}, \textit{self-consistent}, \textit{state-transfer}, and \textit{projected} operators for the orbital rotation and active-space excitation operators. 
\paragraph{Naive operators}
The {naive} orbital rotation operator for singlet excitations is defined as:
\begin{equation}
    \hat{q}_{pq} = \frac{1}{\sqrt{2}}\hat{E}_{pq}
\end{equation}
For the orbital rotation parameterization, the indices are restricted to being in the set $pq\in\left\{ai,av,vi\right\}$.
The naive active-space spin-adapted singlet single and double excitation operators are parametrized according to the following:
\begin{align}
    \hat{G} \in \Bigg\{\frac{1}{\sqrt{2}}\hat{E}_{v_av_i},\quad &\frac{1}{2\sqrt{\left(1+\delta_{v_av_b}\right)\left(1+\delta_{v_iv_j}\right)}}\left(\hat{E}_{v_av_i}\hat{E}_{v_bv_j} + \hat{E}_{v_av_j}\hat{E}_{v_bv_i}\right),\\\nonumber
    &\frac{1}{2\sqrt{3}}\left(\hat{E}_{v_av_i}\hat{E}_{v_bv_j} - \hat{E}_{v_av_j}\hat{E}_{v_bv_i}\right)\Bigg\}.
\end{align}
This parametrization guarantees that only excitations of singlet character will be calculated.~\cite{Paldus1977,Piecuch1989,Packer1996} The following three operator choices are based on transformations of the naive orbital rotation and the naive active-space excitation operator.

\paragraph{Self consistent operators}
The self-consistent (SC) operators~\cite{Asthana2023,Kumar2023,Kim2023} are constructed via a unitary transformation with the wave function active space ansatz $\boldsymbol{U} = \exp\left(-\hat{t}\left(\boldsymbol{\theta}\right)\right)$ (recall \autoref{sec:theo:ooUCC} for our oo-UCC wave function)
\begin{equation}
    \hat{O}^\text{sc} = \boldsymbol{U}\hat{O}\boldsymbol{U}^\dagger~.
\end{equation}
Applying the self-consistent transformation to the orbital rotation operator yields terms of the form:
\begin{equation}
    \hat{q}^\text{sc}_{pq}\left|0\right> = \boldsymbol{U}\hat{q}_{pq}\left|\text{CSF}\right> \label{eq:UqCSF}
\end{equation}
Since the orbital rotation operator now works directly on the reference CSF, the non-redundant parameter space is reduced to $pq\in\left\{v_ai,ai,av_i\right\}$, and the parameters removed are of the type $\left\{v_ii,av_a\right\}$.

\paragraph{State-transfer operators} 
The state-transfer (ST) operators are defined as
\begin{equation}
    \hat{O}^\text{st} = \boldsymbol{U}\hat{O}\left|\text{CSF}\right>\left<0\right|\label{eq:state_transfer_operator}.
\end{equation}
Following the consideration as for the self-consistent orbital rotations, the state-transfer orbital rotation also has the parameter space reduced to $pq\in\left\{v_ai,ai,av_i\right\}$.

\paragraph{Projected} 
Finally, the projected (proj) operators~\cite{Kumar2023} are
\begin{equation}
    \hat{O}^\text{proj} = \hat{O}\left|0\right>\left<0\right| - \left<0\left|\hat{O}\right|0\right>\label{eq:projection_operator}
\end{equation}
and since $\hat{q}^\dagger\left|0\right>=0$, the projection for the orbital rotation reduces to $\hat{q}^\text{proj} = \hat{q}\left|0\right>\left<0\right|$. The transformation of the active-space excitation operators does not lead to any similar reduction in parameter space for any of the transformations.

The combination of the four choices of operators used for the active-space excitation operators $\hat{R}_n$, with the four ones for the orbital rotation, $\hat{Q}_\mu$, makes for sixteen different qLR methods to choose from.
After excluding redundant combinations and combinations that treat orbital rotations at a higher level than the active-space excitations, e.g.\ naive active-space excitations together with state-transfer orbitals rotations, we are left with eight different qLR methods that we investigate in this work.  
These combinations are:
\begin{enumerate}
    \item Naive linear response (\textit{naive LR}),
 using $\hat{R}=\hat{G}$ and $\hat{Q}=\hat{q}$; 
 \item State-transfer linear response (\textit{ST LR}), using $\hat{R}=\boldsymbol{U}\hat{G}\left|\text{CSF}\right>\left<0\right|$ and $\hat{Q}=\hat{q}$; 
 \item Self-consistent linear response
 (\textit{SC LR}), using $\hat{R}=\boldsymbol{U}\hat{G}\boldsymbol{U}^\dagger$ and $\hat{Q}=\hat{q}$; 
 \item Projected linear response (\textit{proj LR}), using $\hat{R}=\hat{G}\left|0\right>\left<0\right| - \left<0\left|\hat{G}\right|0\right>$ and $\hat{Q}=\hat{q}$; 
 \item All state transfer LR (\textit{all-ST LR}), using $\hat{R}=\boldsymbol{U}\hat{G}\left|\text{CSF}\right>\left<0\right|$ and $\hat{Q}=\boldsymbol{U}\hat{q}\left|\text{CSF}\right>\left<0\right|$;
 \item All self-consistent LR
(\textit{all-SC LR}), using $\hat{R}=\boldsymbol{U}\hat{G}\boldsymbol{U}^\dagger$ and $\hat{Q}=\boldsymbol{U}\hat{q}\boldsymbol{U}^\dagger$; \item All projected LR (\textit{all-proj LR}) using $\hat{R}=\hat{G}\left|0\right>\left<0\right| - \left<0\left|\hat{G}\right|0\right>$ and $\hat{Q}=\hat{q}\left|0\right>\left<0\right|$; 
\item State-transfer projected LR (\textit{ST-proj LR}), using $\hat{R}=\boldsymbol{U}\hat{G}\left|\text{CSF}\right>\left<0\right|$ and $\hat{Q}=\hat{q}\left|0\right>\left<0\right|$,
\end{enumerate}
whose characteristics are summarized in \autoref{tab:methods}.
By inserting the definitions of naive, self-consistent, state-transfer, and projected operator in place of the generic active-space excitation operator, $\hat{R}_n$, and of the generic orbital rotation, $\hat{Q}_\mu$ in the eigenvalue equation, \autoref{eq:LR_nopert}, and in the definition of the oscillator strength, \autoref{eq:osc_strength}, we obtain the working 
equations for our eight qLR methods, which are presented in \autoref{sec:LR_WE} and \ref{sec:osc_WE}.

Note that some transformations of this kind have been used in the literature for the full-space equation of motion (EOM)/LR approaches, i.e., approaches without active space and orbital optimization. This includes naive,~\cite{Ollitrault2020,Rizzo2022}self-consistent,~\cite{Asthana2023,Kumar2023,Kim2023} and projected~\cite{Kumar2023} full-space operators. 

\subsection{qLR on near-term quantum hardware\label{sec:qLR_onhardware}}
All qLR methods above can be implemented on a quantum device. However, they differ in the amount and kind of expectation values to be calculated/measured, marking some of them as unsuitable for the near term. Moreover, just as for the oo-VQE approach (\autoref{sec:theo:ooUCC}), the implementation of qLR entails splitting the simulation of operators acting on the active space on quantum devices and the simulation of operators acting outside the active space on classic architecture (recall \autoref{eq:op_splitting}). 

\subsubsection{Expectation values on quantum computers\label{sec:theory_exp_quantum}}
We can differentiate the following kinds of expectation values that appear in the working equations above.
First, in naive and projected operator transformations, we deal with expectation values containing a product of operators, $\hat{O}_i^\text{Pauli} \in \left\{ \hat{q}_\mu, \hat{G}_n, \hat{H}\right\}$, that can be expressed as Pauli strings on a quantum computer, 
\begin{equation}
    M = \braket{0 | \hat{O}_P | 0}\label{eq:simple_measurement}
\end{equation}
with $\hat{O}_P = \prod_i \hat{O}_i^\text{Pauli}$.
These expectation values are obtained by a simple Pauli string measurements on a quantum device.
Second, expectation values that contain additionally the unitary active-space transformation, $\boldsymbol{U}$, are present if U-transformed operators are used, e.g., \ within the self-consistent or state-transfer LR approaches. Here, we make a distinction between diagonal and off-diagonal elements, respectively,
\begin{align}
    M_{l,l}&=\braket{\text{CSF} | \hat{X}_l^\dagger \boldsymbol{U}^\dagger \hat{O} \boldsymbol{U} \hat{X}_l | \text{CSF}}, \\
    M_{l,m}&=\braket{\text{CSF} | \hat{X}_l^\dagger \boldsymbol{U}^\dagger \hat{O} \boldsymbol{U} \hat{X}_m  | \text{CSF}}, \label{eq:expU:off}
\end{align}
with $ \hat{X}_l \in \left\{ \hat{q}_\mu, \hat{G}_n \right\}$. The diagonal element, $M_{l,l}$, can be measured directly as it follows the structure of \autoref{eq:simple_measurement} with $\left|0\right>=\boldsymbol{U} \hat{X}_l\left|\text{CSF}\right>$.
The off-diagonal element, $M_{l,m}$, needs to be evaluated using a Hadamard-test circuit, an expensive algorithm with deep circuits and ancilla qubits that is not classified near-term.\cite{aharonov2006polynomial,knill2007optimal,dobvsivcek2007arbitrary}

For the specific case of Hermitian operators, $\hat{O} = \hat{O}_H$, it has been shown~\cite{Parrish2019,Nakanishi2019} that the off-diagonal expectation values in \autoref{eq:expU:off} can be calculated as
\begin{align}
    M_{l,m} &= \left<\text{CSF}\left|\hat{X}_l^\dagger\boldsymbol{U}^\dagger\hat{O}_\text{H}\boldsymbol{U}\hat{X}_m\right|\text{CSF}\right> \\
    \Re\left[M_{l,m}\right] &= \frac{1}{2}\left(\left<\text{CSF}\left|\left(\hat{X}_l^\dagger + \hat{X}_m^\dagger\right)\boldsymbol{U}^\dagger\hat{O}_\text{H}\boldsymbol{U}\left(\hat{X}_l + \hat{X}_m\right)\right|\text{CSF}\right>\right.\label{eq:kumar_trick}\\\nonumber
    &\quad\left. - \left<\text{CSF}\left|\hat{X}_l^\dagger\boldsymbol{U}^\dagger\hat{O}_\text{H}\boldsymbol{U}\hat{X}_l\right|\text{CSF}\right> - \left<\text{CSF}\left|\hat{X}_m^\dagger\boldsymbol{U}^\dagger\hat{O}_\text{H}\boldsymbol{U}\hat{X}_m\right|\text{CSF}\right>\right),
\end{align}
which only utilizes expectation values that can be calculated according to \autoref{eq:simple_measurement} without Hadamard test circuits. 

\subsubsection{Approximate Hermitification\label{sec:herm}}

Both the self-consistent formulation and the state-transfer formulation (see working equations in the appendix, \autoref{eq:sc_A} and \ref{eq:sc_B}, and \autoref{eq:st_A} and \ref{eq:st_B}, respectively), have off-diagonal elements of the type:
\begin{equation}
    \boldsymbol{A}^{Gq} = \left<\text{CSF}\left|\hat{G}_n^\dagger\boldsymbol{U}^\dagger\hat{H}\hat{q}_\mu\right|0\right>
\end{equation}
and,
\begin{equation}
    \boldsymbol{B}^{Gq} = -\left<\text{CSF}\left|\hat{G}_n^\dagger\boldsymbol{U}^\dagger\hat{q}_\mu^\dagger\hat{H}\right|0\right>
\end{equation}

Since the operators $\hat{H}\hat{q}_\mu$ and $\hat{q}_\mu^\dagger\hat{H}$ are non-Hermitian, the decomposition in \autoref{eq:kumar_trick} cannot be applied to calculate these elements and a Hadamard-test circuit is needed.
In an attempt to alleviate the problem, we present an approximate Hermitification. Consider what happens to the $\boldsymbol{B}$ matrix element in the limit of the wave function, $\left|0\right>$, being the exact solution:
\begin{align}
    \nonumber
    \boldsymbol{B}^{Gq} &= -\left<\text{CSF}\left|\hat{G}_n^\dagger\boldsymbol{U}^\dagger\hat{q}_\mu^\dagger\hat{H}\right|\text{FCI}\right>\\
    \nonumber
    &= -\left<\text{CSF}\left|\hat{G}_n^\dagger\boldsymbol{U}^\dagger\hat{q}_\mu^\dagger\right|\text{FCI}\right>E_0\\ 
    &= 0
\end{align}
Here it has been used that $\hat{H}\left|\text{FCI}\right> = E_0\left|\text{FCI}\right>$ and $\hat{q}_\mu^\dagger\left|0\right> = 0$.
For any wave function that is a good approximation to the FCI solution, the $\boldsymbol{B}$ matrix elements are close to zero.
Now consider the $\boldsymbol{A}$ matrix element, where we subtract zero ($-\boldsymbol{B}^{Gq}$):
\begin{align}
\nonumber
    \boldsymbol{A}^{Gq} &\approx \left<\text{CSF}\left|\hat{G}_n^\dagger\boldsymbol{U}^\dagger\hat{H}\hat{q}_\mu\right|0\right> + \left<\text{CSF}\left|\hat{G}_n^\dagger\boldsymbol{U}^\dagger\hat{q}_\mu^\dagger\hat{H}\right|0\right>\\
    &= \left<\text{CSF}\left|\hat{G}_n^\dagger\boldsymbol{U}^\dagger\left(\hat{H}\hat{q}_\mu+\hat{q}_\mu^\dagger\hat{H}\right)\right|0\right>
\end{align}
Since the operator $\hat{H}\hat{q}_\mu+\hat{q}_\mu^\dagger\hat{H}$ is Hermitian, \autoref{eq:kumar_trick} can be applied to calculate the elements of the $\boldsymbol{A}$ matrix.
This approximation to calculate the $\boldsymbol{A}$ matrix elements only holds for wave functions where $\boldsymbol{B}^{Gq}\approx 0$.

\subsubsection{Feasibility of qLR methods}
\autoref{tab:methods} summarizes the number of unique expectation values to be measured on a quantum device for each qLR method and classifies its feasibility for near-term application. The number of terms is obtained from the respective working equations in \autoref{sec:LR_WE}, as the number of unique expectation values. Near-term reflects that the method's expectation values can be evaluated by simple Pauli strings measurements without expensive Hadamard test circuits (cf. \autoref{sec:theory_exp_quantum}).  
\begin{table}[htbp!]
\caption{Overview over qLR methods with their operator transformations, amount of generic terms for the eigenvalue problem evaluation of excited states, and feasibility for implementation on quantum hardware. nt = near-term, nt+decomp = near-term using \autoref{eq:kumar_trick} decomposition, htc: needs Hadamard test circuit, herm = terms needed using approximated hermitification (\autoref{sec:herm}), decomp: terms needed using \autoref{eq:kumar_trick} decomposition.}
\begin{center}
\begin{tabular}{l|ccc|cc}
\textbf{qLR} & $\boldsymbol{\hat{R}}$ & $\boldsymbol{\hat{Q}}$ & \textbf{generic terms} & \multicolumn{ 2}{c}{\textbf{feasibility}} \\ \hline
naive & $\hat{G}$ & $\hat{q}$ & 18 & nt & - \\ 
SC & $\boldsymbol{U}\hat{G}\boldsymbol{U}^\dagger$ & $\hat{q}$ & 9 & htc & herm: 16 terms \\ 
ST & $\boldsymbol{U}\hat{G}\left|\text{CSF}\right>\left<0\right|$ & $\hat{q}$ & 7 & htc & herm: 10 terms \\ 
proj & $\hat{G}\left|0\right>\left<0\right| - \left<0\left|\hat{G}\right|0\right>$ & $\hat{q}$ & 10 & nt & - \\ 
all-SC & $\boldsymbol{U}\hat{G}\boldsymbol{U}^\dagger$ & $\boldsymbol{U}\hat{q}\boldsymbol{U}^\dagger$ & 8 & nt+decomp &  decomp: 24 terms \\ 
all-ST & $\boldsymbol{U}\hat{G}\left|\text{CSF}\right>\left<0\right|$ & $\boldsymbol{U}\hat{q}\left|\text{CSF}\right>\left<0\right|$ & 3 & nt+decomp & decomp: 9 terms \\ 
all-proj & $\hat{G}\left|0\right>\left<0\right| - \left<0\left|\hat{G}\right|0\right>$ & $\hat{q}\left|0\right>\left<0\right|$ & 7 & nt & - \\ 
ST-proj & $\boldsymbol{U}\hat{G}\left|\text{CSF}\right>\left<0\right|$ & $\hat{q}\left|0\right>\left<0\right|$ & 4 & htc & herm: 8 terms \\ \hline
\end{tabular}
\end{center}
\label{tab:methods}
\end{table}
For near-term (nt) application, naive LR, proj LR, and all-proj LR are feasible and necessitate the simulation of 18, 10, and 7 unique expectation value terms for the eigenvalue problem (\autoref{eq:LR_nopert}), respectively. 
Using the decomposition approach 
in \autoref{eq:kumar_trick}, 
all-SC and all-ST can be used near-term. The decomposition leads to an increase in terms by a factor of three; thus, 24 and 9 unique terms are needed, respectively. 
The methods SC LR, ST LR, and ST-proj LR need Hadamard test circuits (htc), which classifies these methods as not near-term. However, as elaborated in \autoref{sec:herm}, we propose an approximated Hermitification that brings these methods to near term with 16, 10, and 8 unique terms to be evaluated, respectively. This analysis shows that for near-term quantum hardware, especially the all-proj LR, all-ST LR and proj LR are attractive methods with a minimal number of 
unique terms. Beyond the near-term, i.e., \ with a quantum computer capable of Hadamard test circuits, ST-proj LR and all-ST LR will be interesting candidates with only four and three unique terms, respectively. 

Lastly, we want to point out that to solve the eigenvalue problem in \autoref{eq:LR_nopert}, the metric, $\boldsymbol{S}^{[2]}$, has to be inverted. For small values within the metric, noise in the expectation value simulations can become a bottleneck of the algorithm. Investigations of this will be part of future works. 

\section{\label{sec:comp}Computational Details}
In the following, we will adopt the CAS-like active-space notation WF($n$, $o$), where $n$ is the number of electrons in the active space and $o$ is the number of spatial orbitals in the active space.
It is implied that all inactive orbitals are doubly occupied and all virtual orbitals are unoccupied.
All simulations utilized basis set exchange\cite{Pritchard2019,Feller1996,Schuchardt2007} for the 
6-31G~\cite{Dill1975,Ditchfield1971,Hehre1972} and cc-pVDZ~\cite{Dunning1989,Prascher2010} basis set.

The oo-UCC($n$, $o$) calculations were performed using our in-house developed quantum chemistry oo-UCC code SlowQuant~\cite{slowquant} interfaced with PySCF~\cite{Sun2015,Sun2020} for the one-electron and two-electron atomic orbital integrals, as well as for the initial MP2 natural orbitals.~\cite{Mller1934,Jensen1988} The wave function optimization was performed by doing a UCC($n$, $o$) to get the UCC amplitudes to match the MP2 natural orbitals. Afterward, the oo-UCC($n$, $o$) calculation was performed, except for the one at the equilibrium structure of BeH$_2$, which was started from Hartree-Fock orbitals.

The classic CASSCF($n$, $o$) and FCI calculations were performed with the Dalton Project~\cite{Olsen2020} using
Dalton~\cite{Aidas2013,DALTON2022} as a backend.

A geometry of H$_4$  at the FCI $S_0$/$S_1$ minimum-energy conical intersection (MECI) point (in the 6-31G basis) was optimized using the GeomeTRIC program\cite{wang2016geometry}, with energies and gradients of the ground- and first excited state computed using the Dalton program. Since no Dalton engine is available by default, we used the custom engine interface available in GeomeTRIC. 
The MECI optimization in GeomeTRIC implements the penalty-constrained approach described by \citeauthor{levine2008optimizing},\cite{levine2008optimizing}, which does not require non-adiabatic coupling vectors. The MECI was optimized with parameters $\sigma=5.0$ and $\alpha=0.01$.
We note that this geometry is a MECI only with FCI/6-31G and that it will only be in the neighborhood of such a point for the other tested methods.

\section{\label{sec:results}
Results and Discussion}
Having derived new working equations for qLR in active spaces (see \autoref{sec:LR_WE} and \ref{sec:osc_WE}) and analyzed their near-term feasibility, we now contrast our qLR methods against classic quantum chemistry methods. We aim to test the various operator transformations and shed light on the effect of the truncation in the excitation space of qLR. 

We simulate excited state potential energy curves (PECs) and absorption spectra with our various qLR approaches to a level of singles and doubles excitation, i.e., \ LRSD. We compare the results to classic CASSCF LR and FCI LR simulations. 
As CASSCF performs a complete CI simulation within the active space, a performative criterion for our methods is how they compare to their classic complete space counterparts.

For ease of comparison, we group, at certain points, our qLR methods by the transformation that is applied to the orbital rotations. Specifically, \emph{naive-q LR} groups together naive LR, SC LR, ST LR, and proj LR; \emph{proj-q LR} groups together all-proj LR and ST-proj LR, while \emph{U-q LR} groups together all-SC LR and all-ST LR. 
We apply this grouping when the methods within a certain group all give identical results. 

\subsection{Excited state potential energy curves}

\subsubsection{LiH \textit{(2,2)}\label{sec:results_LiH}}

\begin{figure}[tbp]
	\centering 
	\includegraphics[width=0.7\textwidth]{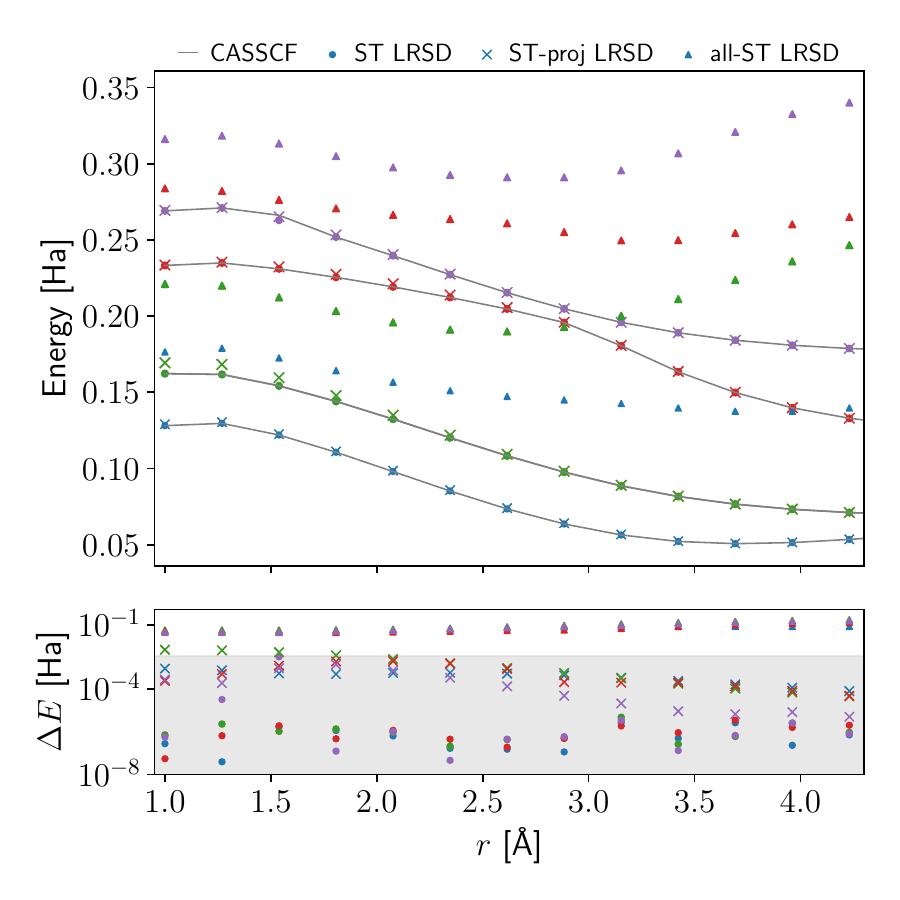}
	\caption{Top: first four excited states (blue, green, red, purple) of LiH \textit{(2,2)} / cc-pVDZ obtained with selected qLR methods all using ST for the active-space excitations but different orbital rotation parametrization, namely naive-q (ST LRSD, dots), proj-q (ST-proj LRSD, crosses), and U-q (all-ST LRSD, triangles). Comparisons with CASSCF (gray solid line). Bottom: Energy differences between qLR methods and CASSCF. Gray background indicates chemical accuracy.}
	\label{fig:LiH_631G}
\end{figure}

In \autoref{fig:LiH_631G}, the first four singlet excitation energies of LiH \textit{(2,2)} for selected methods can be seen as a function of the internuclear separation.
The active space parametrization of the qLR is state-transfer for all of the cases.
Choosing another active-space parameterization gives the exact same results because the system has only two electrons in two orbitals, and hence, all of the active-space parameterizations span the same excitation space.
The difference is in the parametrization of the orbital rotations: ST LRSD uses naive $q$, ST-proj LRSD uses projected $q$, and all-ST LRSD uses state-transfer $q$.
The ST LRSD with naive $q$ (dots) gives exactly the same excitation energies (within convergence) as the CASSCF reference.
This was expected since the response 
parametrization of the two methods is identical, and the wave function is, in principle, also identical as for an active space of \textit{(2,2)}, the truncated LRSD and the complete CI active space in CASSCF coincide. 

When comparing the excitation energies of the ST-parametrized orbital rotations (all-ST LRSD, triangles) with the excitation energies from CASSCF, it can be seen that excitation energies from the 
ST-parametrized orbital rotations have a significant error, up to 0.1 eV, with respect to the CASSCF excitation energies.
The difference between the excitation energies from the ST-parametri\-zed orbital rotations and the CASSCF increases with the distance between the lithium and hydrogen atoms.
This indicates that the more this orbital rotation parameterization breaks down, the more multi-configurational the molecular system becomes.
The reason for the breakdown of 
ST-parametrized orbital rotation was discussed in \autoref{sec:transform} when discussing \autoref{eq:UqCSF}. 
Namely, the unitary transformation in the ST (and SC) formalism leads to the orbital rotations acting on the reference CSF instead of the multi-configurational $\ket{0}$. This leads to an 
under-parametrization of the LR equations and strong deviations in excitation energies. 
As ST and SC parameterized orbital rotation is based on the same formalism, all-ST LR and all-SC LR suffer from the same under-parametrization. Thus, both methods are not analyzed further in the following. 

The comparison of projected parametrized orbital rotations (ST-proj LRSD, crosses) with CAS\-SCF reveals small differences, mostly within chemical accuracy. These differences originate from the fact that projected orbital rotations are fundamentally novel methods compared to traditional (classic) naive orbital rotations. In \autoref{sec:results_spectra}, we will go into more detail on this and show that this novel parametrization leads to different results compared to classic CASSCF with naive $q$, yet comparing to FCI simulations will reveal that these are not worse results, but rather results of a new class of methods. 

\subsubsection{H$_4$ \textit{(4,4)}\label{sec:results_H4}}
\begin{figure}[htbp!]
	\centering 
\includegraphics[width=.7\textwidth]{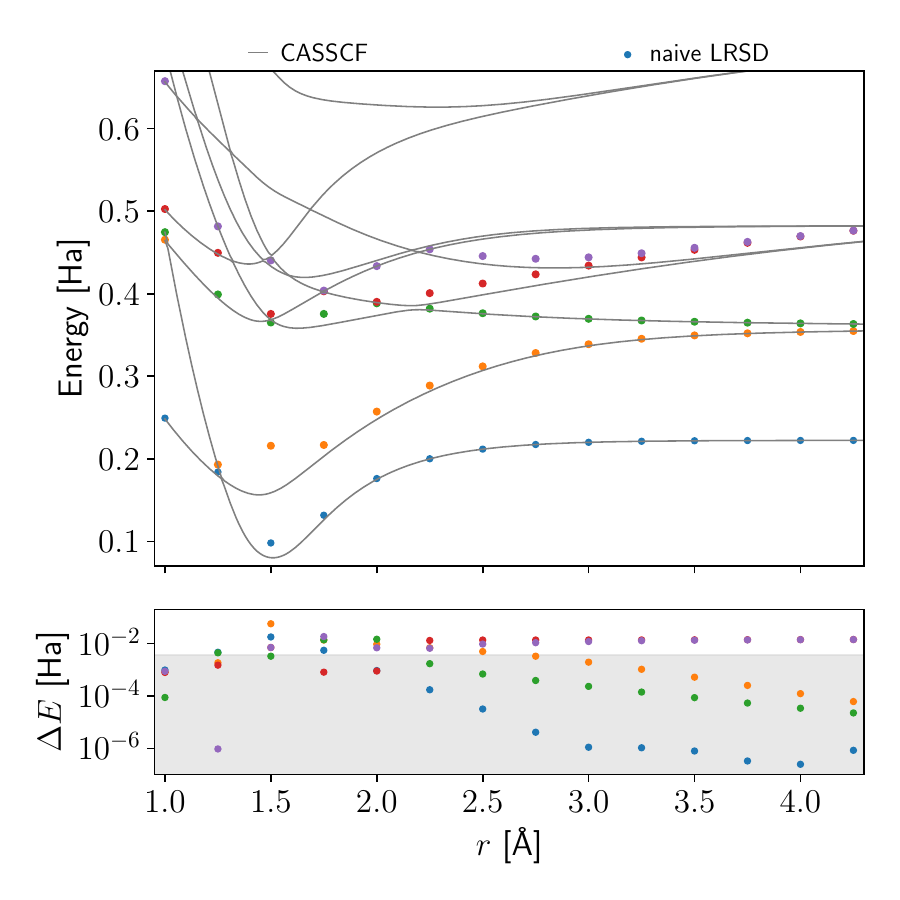}
	\caption{Top: Excited state PEC of the first five states (blue, orange, green, red, purple, brown) for rectangular H$_4$ \textit{(4,4)} / cc-pVDZ. Naive LRSD (dots) results are contrasted to CASSCF (solid gray lines). Bottom: Energy differences of naive LRSD to CASSCF. The gray background indicates chemical accuracy.}
	\label{fig:H4_naive}
\end{figure}
The rectangular H$_4$ molecule has been proven to be a difficult test case for quantum excited-state methods in full space with minimal basis and using the naive parametrization in the active space.\cite{Asthana2023,Kumar2023}
In \autoref{fig:H4_naive}, we present our results for an active space of \textit{(4,4)} 
with the cc-pVDZ basis set for naive LRSD, while the results for all other qLR methods are presented in the appendix (see \autoref{fig:H4_all}). The distance $r$ between two parallel H$_2$ molecules (each with a fixed bond length of $1.5\,$\AA{}) is scanned as the PEC coordinate. \autoref{fig:H4_naive} shows how naive LR cannot reproduce the excited state behavior at long bond lengths, with an error at the order of $10^{-2}\,$Ha compared to CASSCF. Moreover, all methods exhibit deviations around the highly degenerate point of $r = 1.5\,$\AA{} (square H$_4$), with naive LR having the largest deviations. This is consistent with previous work\cite{Asthana2023,Kumar2023} and shows that, even though naive LR has the most generic terms to be simulated, it does not reproduce the correct energies for certain systems when using a truncation to singles and doubles in the active space. 

\subsection{Absorption spectra \label{sec:results_spectra}}
For the simulation of absorption spectra, we focus on the qLR methods using naive orbital rotations, i.e., \ the \emph{naive-q LR} grouping: naive LR, SC LR, ST LR, and proj-LR, and qLR methods using projected orbital rotations, i.e.,\ the \emph{proj-q LR} grouping: all-proj LR and ST-proj LR. The methods all-SC LR and all-ST LR will not be discussed any further since they were proven erroneous in 
\autoref{sec:results_LiH}.

\subsubsection{N$_2$ \textit{(6,6)}}

\begin{figure}[htbp!]
	\centering 
	\includegraphics[width=0.7\textwidth]{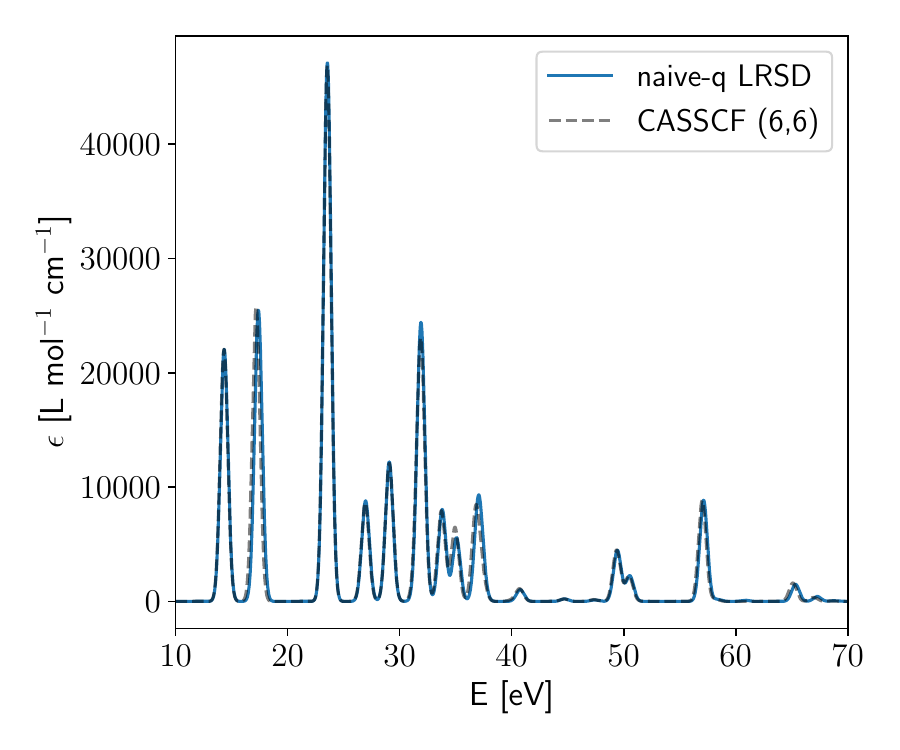}
	\caption{Comparison of the absorption spectrum of N$_2$ \textit{(6,6)} / 6-31G from 
 naive-q LRSD (blue line) with the one from CASSCF (dashed gray line).
 }
	\label{fig:N2}
\end{figure}

As mentioned before, the classic CASSCF method utilizes naive orbital rotations and, thus, resembles our naive-q LR methods. However, as we perform a truncation in the active space to the singles and doubles level, we dramatically reduce the number of active-space parameters. In the case of N$_2$~\textit{(6,6)}, this leads to a reduction from 174 to 54 configurations compared to CASSCF. Nonetheless, \autoref{fig:N2} shows that there is no significant change in the absorption spectrum as a consequence of this truncation. Moreover, the same results are obtained for all naive-q methods, regardless of their active-space parametrization. 

Therefore, using our qLR approaches with truncation at the singles and doubles level allows 
to dramatically reduce
the size of the Hessian and operator pool for the linear response equations whilst retaining an accuracy comparable to the complete active-space level in CASSCF.

\subsubsection{H$_2$O \textit{(4,4)}\label{sec:results_H2O}}
\begin{figure}[htbp!]
	\centering 
	\includegraphics[width=0.7\textwidth]{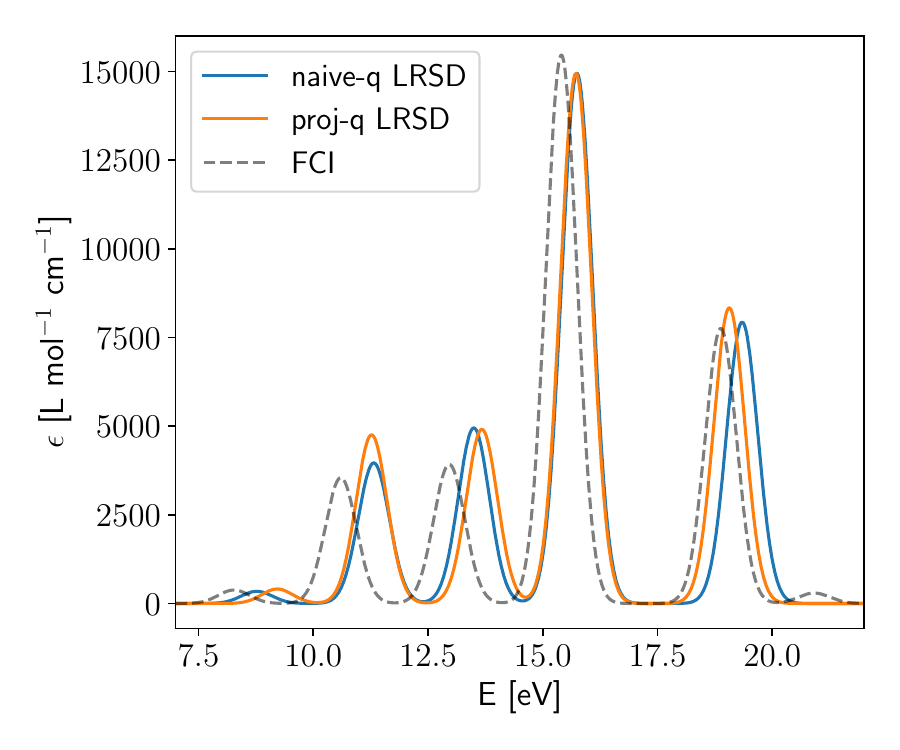}
	\caption{Comparison of the absorption spectrum of H$_2$O \textit{(4,4)} / 6-31G obtained with naive-q LRSD (solid blue line), proj-q LRSD (solid orange line), and FCI (dashed gray line).}
	\label{fig:H2O_eq_631G_FCI_zoomed}
\end{figure}

Analysis of the absorption spectra of H$_2$O with a (4,4) active space yields the same conclusion as in the N$_2$ example above. Comparing naive-q LRSD with CASSCF results in a reduction from 19 to 14 active-space excitation parameters that do not influence the accuracy of the spectrum significantly (see appendix, \autoref{fig:H2O_eq_631G_fullrange}).
Here, we want to focus on the spectrum obtained with proj-q LRSD methods, i.e., \ all-proj LRSD and ST-proj LRSD that transform the orbital rotation parameters via a projection. These spectra differ from the naive-q LRSD/CASSCF results (see appendix \autoref{fig:H2O_eq_631G_fullrange}). However, as \autoref{fig:H2O_eq_631G_FCI_zoomed} reveals, they do not differ for the worse. The comparison with the spectrum obtained by FCI LR shows that naive-q LRSD/CASSCF and proj-q LRSD differ from the FCI results in a way that it is unclear which method is better. Hence, using proj-q LR methods can be understood as a new class of qLR methods (compared to naive-q LR) that give different, yet no worse, results due to their different parametrization of the orbital rotations. 

\subsubsection{BeH$_2$ \textit{(4,6)} at equilibrium and stretched geometries\label{sec:results_BeH2}}

\begin{figure}[htbp]
	\centering 
    \subfigure{
	\includegraphics[width=.5\textwidth]{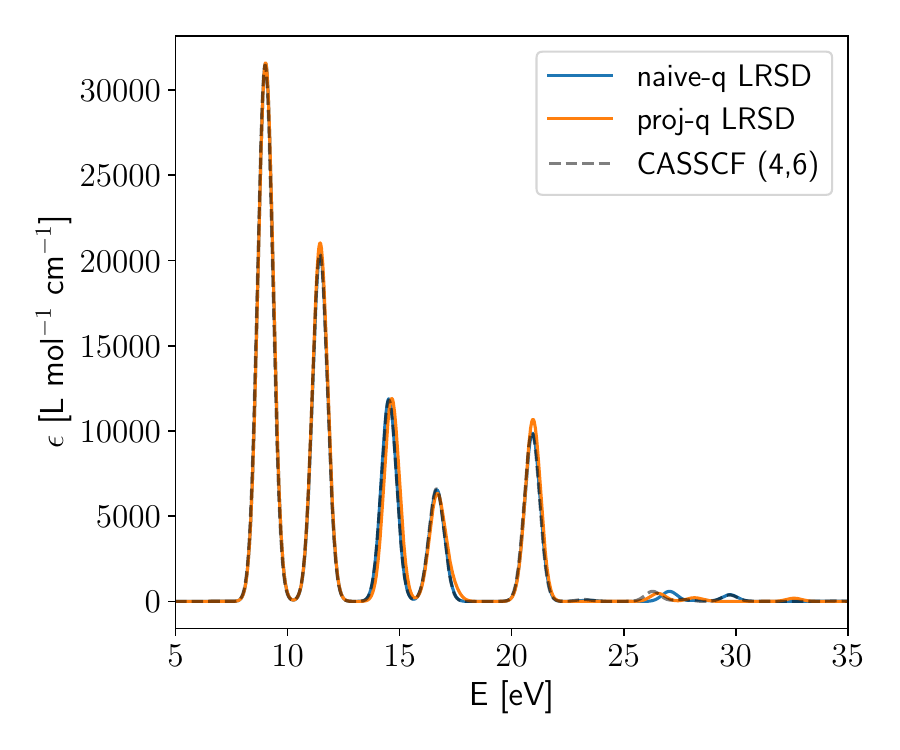}}%
    \subfigure{
    \includegraphics[width=.5\textwidth]{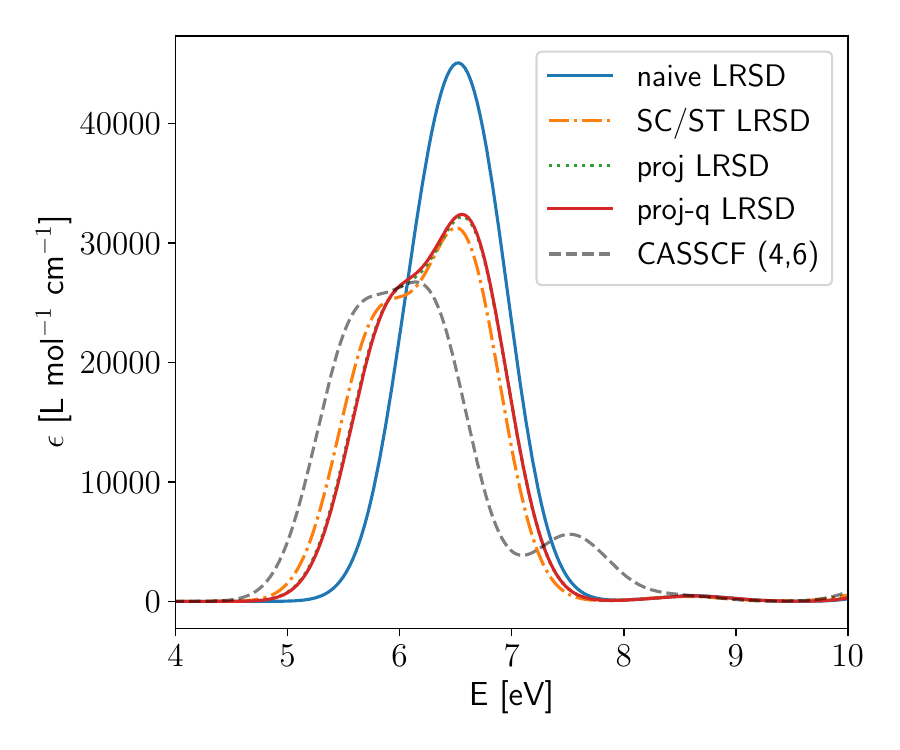}}
	\caption{Absorption spectra of BeH$_2$ \textit{(4,6)} / 6-31G. Left: At equilibrium geometry, comparison of naive-q LRSD methods (solid blue line) and proj-q LRSD methods (solid orange line) to CASSCF (gray dashed line). Right: At symmetrically stretched (doubled bond length) geometry, comparison of various qLR methods to CASSCF \textit{(4,6)} in a selected (excitation) energy range.}
	\label{fig:BeH2_46_631G_CAS}
\end{figure}

For BeH$_2$ \textit{(4,6)}, we observe a reduction in active-space parameters from 104 to 44. \autoref{fig:BeH2_46_631G_CAS} (left) shows the absorption spectrum for BeH$_2$ in its equilibrium geometry, where naive-q LRSD methods, as well as proj-q LRSD methods, matches CASSCF. In the appendix (see \autoref{fig:BeH2_eq_46_631G_FCI}), the comparison with FCI is shown.

The spectrum of stretched BeH$_2$ \textit{(4,6)} for a restricted energy range is shown in the right panel of \autoref{fig:BeH2_46_631G_CAS} (we refer to \autoref{fig:BeH2_str_46_631G_CAS_full} and \ref{fig:BeH2_str_46_631G_FCI} in the appendix for the full-range spectrum and comparison with FCI comparison) and reveals differences within the qLR methods and compared to CASSCF. As seen for H$_4$ in \autoref{sec:results_H4}, stretched BeH$_2$ \textit{(4,6)} is a highly degenerate example where the reduction in active-space excitation for the qLR methods leads to deviation compared to the complete CI active-space expansion in classic CASSCF. Crucially, the different parametrizations of our qLR methods are impacted by the truncation in different ways. It is evident from \autoref{fig:BeH2_46_631G_CAS} (right) that naive LRSD deviates strongest from the other qLR methods and CASSCF. Not only is the shoulder structure of the low-energy peak missing, but the double-peak structure is also lost. This adds to the previously discussed full-space results in the literature about the downsides of the naive parametrization within the qLR formalism\cite{Asthana2023,Kumar2023}. 

\begin{figure}[b!]
	\centering 
        \subfigure{
	\includegraphics[width=.5\textwidth]{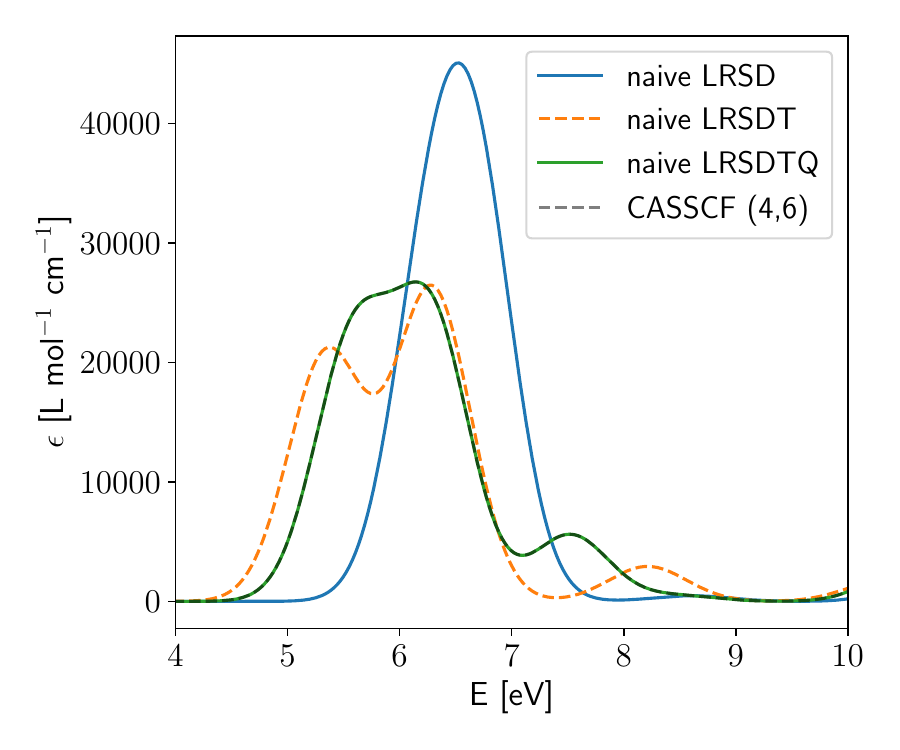}}%
        \subfigure{
	\includegraphics[width=.5\textwidth]{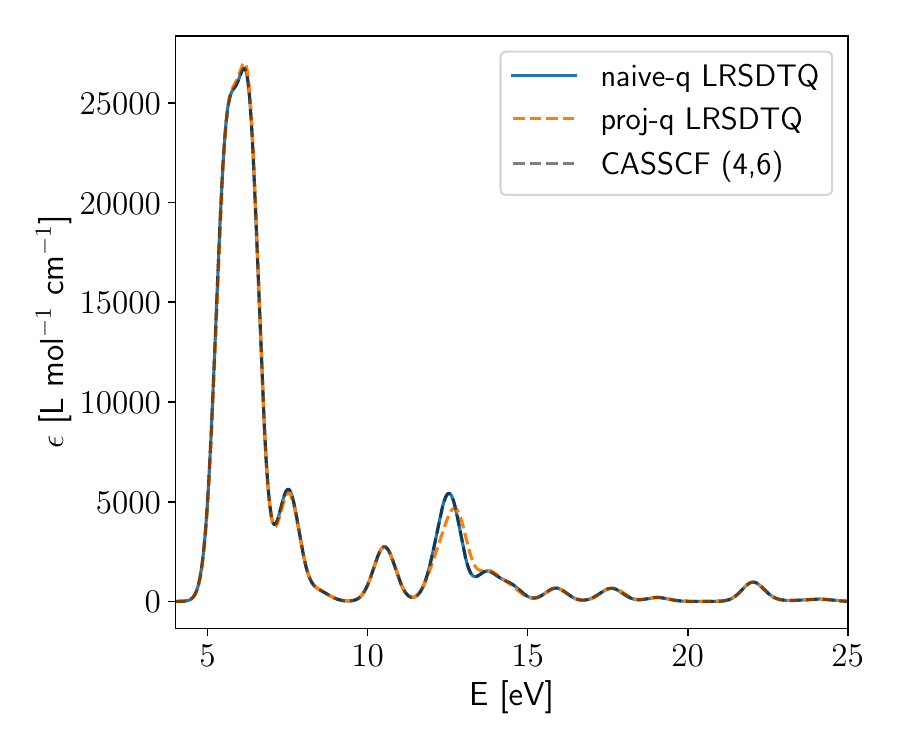}}
	\caption{Absorption spectrum of symmetrically stretched BeH$_2$ \textit{(4,6)} / 6-31G. Left: convergence of naive LR towards the CASSCF results (grey dashed) when varying the rank of excitations in the active space (SD: solid blue, SDT: dashed orange, SDTQ: solid green). The naive LRSDTQ and CASSCF results overlap. Right: Comparison of naive-q LRSDTQ methods (solid blue) and proj-q LRSDTQ methods (dashed orange) to CASSCF (dashed grey). The naive-q LRSDTQ and CASSCF results are overlapping.}
	\label{fig:BeH2_str_46_conv}
\end{figure}
To analyze these deviations from CASSCF further, we performed simulations with larger active-space excitation ranks, namely LRSDT and LRSDTQ, the parameterization of the wave function is also increased, oo-UCCSDT and oo-UCCSDTQ respectively. The latter corresponds to complete CI in the active space for the \textit{(4,6)} active space of stretched BeH$_2$. \autoref{fig:BeH2_str_46_conv} (left) shows the result for naive LR. 
Clearly, the spectrum converges towards the CASSCF results. This is the case for all qLR methods with naive orbital rotations. Hence, the deviations of naive LR from the other qLR methods are due to a slower convergence to the correct, complete CI active-space expansion results of CASSCF. 
However, this does not make naive LR inherently unsuitable for future quantum implementations. Methods using projected orbital rotations (i.e., \ all-proj LR and ST-proj LR) converge to spectra that are very similar to the CASSCF ones but differ in some areas (\autoref{fig:BeH2_str_46_conv} (right)). This reinforces our previous conclusion in \autoref{sec:results_H2O} as proj-q LR methods are a different class of active-space methods due to using a different transformation of the orbital rotations compared to classic 
CASSCF.

\subsection{Hermitification}
\begin{figure}[tbp]
    \centering
    \begin{minipage}{0.49\textwidth}
        \centering
        H$_2$O \textit{(4,4)} / 6-31G\\
        $||B^{Gq}|| = 0.3763$
        \includegraphics[width=\linewidth]{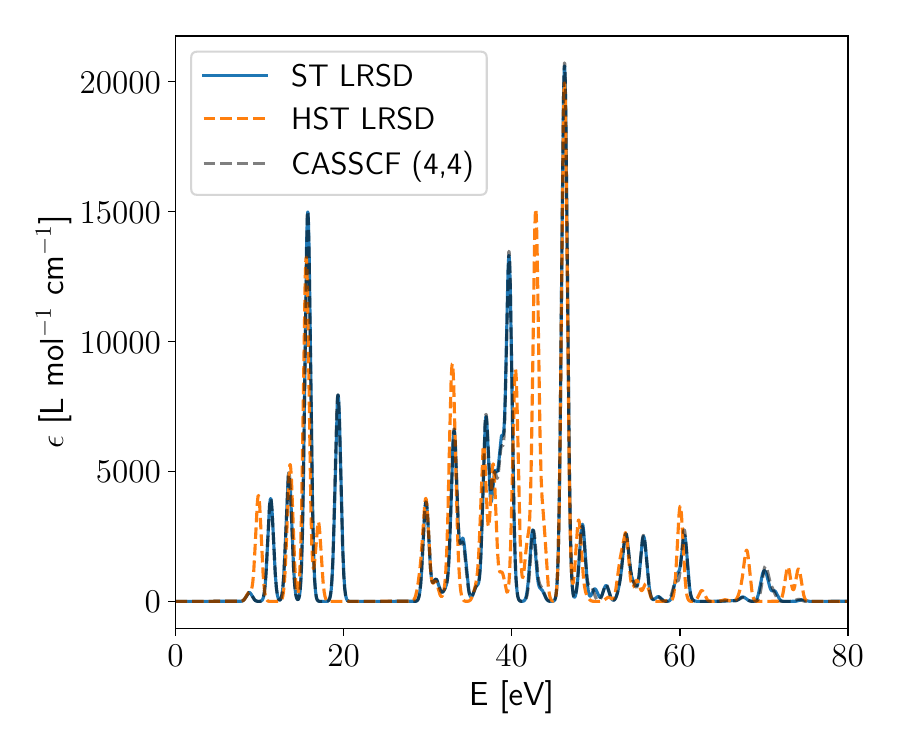}
    \end{minipage}
    \hfill
    \begin{minipage}{0.49\textwidth}
        \centering
        BeH$_2$ \textit{(4,6)} / 6-31G\\
        $||B^{Gq}|| = 0.0708$
        \includegraphics[width=\linewidth]{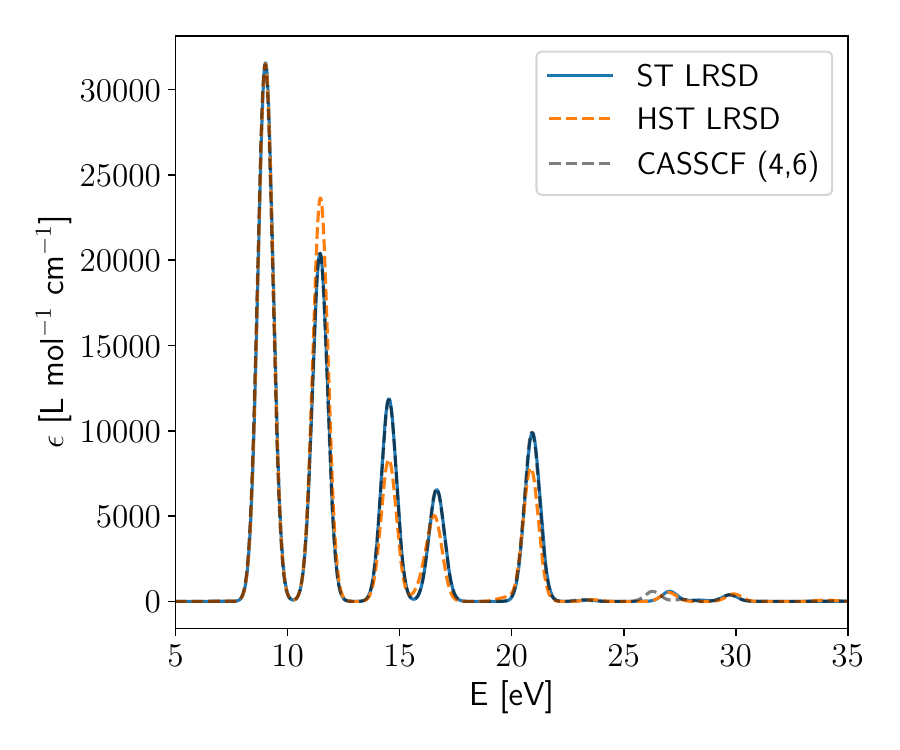} 
    \end{minipage}
    \caption{Examples for the effect of approximate Hermitification using ST LRSD. Solid blue lines are the ST LRSD results, while dashed orange lines are the hermitified ST LRSD results, dubbed HST LRSD.}
    \label{fig:HST}
\end{figure}

In \autoref{sec:herm}, we discussed that ST LR, SC LR, and ST-proj LR are no near-term methods due to the need for Hadamard test circuits to evaluate the matrix elements, and proposed an approximate Hermitification as a possible way to overcome 
the problem. 
As an example, \autoref{fig:HST} shows the impact of this Hermitification on the ST LRSD results, named HST LRSD. We remind the reader that the Hermitification approach assumes that $B^{Gq} \approx 0$. A clear conclusion from \autoref{fig:HST} is that the larger the norm ($||B^{Gq}||$) of the omitted matrix elements is, the larger is the deviation of the HST LRSD spectrum from its ST LRSD counterpart. Therefore, the differences are pronounced for H$_2$O (4,4) / 6-31G with $||B^{Gq}|| = 0.3763$, while they are small for BeH$_2$ (4,6) / 6-31G with $||B^{Gq}|| = 0.0708$. In conclusion, the qLR methods SC/ST/ST-proj LR can be made near-term using Hermitification, but the spectrum quality will depend largely on the (approximate) validity of the condition $B^{Gq} \approx 0$. With better, non-approximated methods at hand, this might not be the near-term way forward.

\subsection{Conical intersection geometry of H$_4$}
\begin{figure}[htbp]
	\centering 
	\includegraphics[width=.7\textwidth]{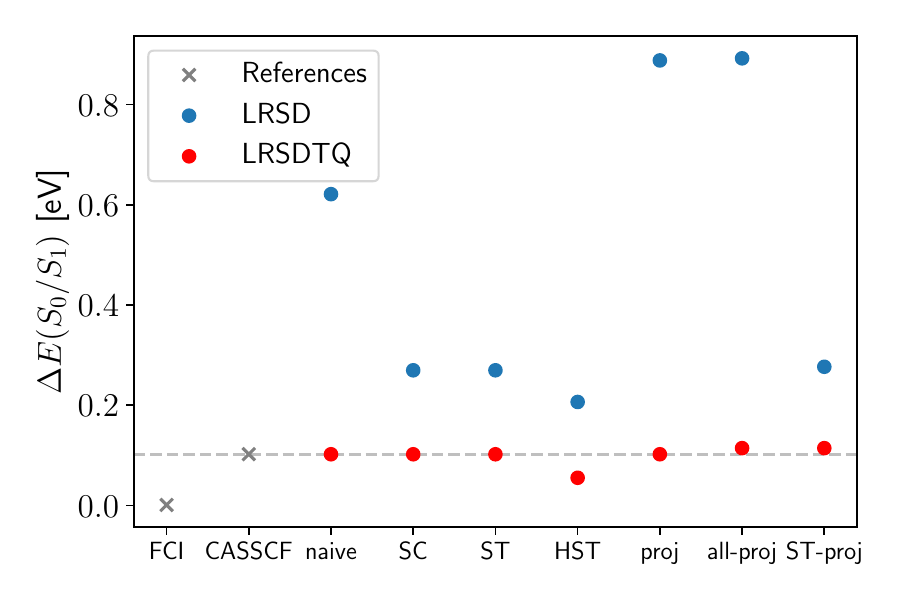}
	\caption{Energy gap between the ground state, $S_0$, and the first excited state, $S_1$, of H$_4$ \textit{(4,4)} / \mbox{6-31G} at the optimized CI. The references FCI and CASSCF (gray crosses), the qLR methods within LRSD (blue dots), and the qLR methods within LRSDTQ (red dots) are contrasted.}
	\label{fig:CI}
\end{figure}

Conical intersections (CIs) are points of degeneracy of at least two potential energy surfaces that drive photochemical processes. Their description poses a great challenge for quantum chemical simulations. Therefore, we chose to test our qLR methods at the H$_4$ ground state CI as a challenging test case to emphasize differences in the parametrizations. \autoref{fig:CI} shows the energy gap between the ground state, $S_0$, and the first excited state, $S_1$. The geometry is a minimum-energy CI point with FCI, so the energy gap is $\Delta E = 0$ for that method. Within LRSD (blue dots), the various qLR methods differ from each other and do not reproduce the CASSCF \textit{(4,4)} reference. Interestingly, SC, ST, and ST-proj LRSD give similar energy gaps, while naive, proj, and all-proj LRSD differ with much larger gaps. In the limit of complete active-space, i.e., \ LRSDTQ, all qLR methods using naive orbital rotation (\emph{naive-q LR}) coincide with CASSCF, as expected from \autoref{sec:results_BeH2}. The proj-q methods (\emph{proj-q LR}) all-proj and ST-proj LRSDTQ give the same result but show slight deviations from CASSCF. This reinforces our previous conclusion of proj-q methods as a new class of method compared to classic CASSCF with naive orbital rotations.
The approximated method HST LRSDTQ shows a smaller error in the calculated excitation energy with respect to FCI than the excitation energy calculated with CASSCF.
This is not due to HST being a good approximation in this case, but rather that some of the $B^\text{Gq}$ elements are significantly different from zero, $||B^{Gq}||=0.12$.
The error introduced just happened to lower the excitation energy.
In \autoref{fig:CI}, it can be seen that the error of the excitation energy calculated with CASSCF with respect to the excitation energy calculated using FCI is about 0.1 eV.
This emphasizes that the electronic structure at the geometry of CIs is very challenging to describe accurately.
In future work, we will look in more detail at how the various qLR methods can be used to find the geometries of CIs and how these results compare. 

\newpage
\section{\label{sec:summary}Summary}
In this paper we have presented and discussed novel formulations of time-dependent linear response theory in active spaces suitable for near-term quantum computers. This theory allows the simulation of excited states and their properties using minimal resources due to the active space approximation, truncation of the active space excitations, and new LR operator transformations. Using a unitary coupled-cluster framework, we present results going beyond the often presented minimal basis sets in quantum computing and show results without significant loss of accuracy compared to classic CASSCF LR theory.

% our approach
Linear response with active spaces is formulated by means of orbital rotation and active-space excitation operators. The latter encompasses a complete CI expansion in classic quantum chemistry theory, such as CASSCF LR. We started out by truncating the active space excitations to the levels of singles and doubles while introducing new ways of representing and transforming the orbital rotation and active-space excitation operators. Therein, we investigated naive, self-consistent, state-transfer, and projected manifolds of operators. Overall, we derived and examined eight different qLR methods, each with a unique operator transformation combination for orbital rotations and active-space excitations. Crucially, the kind of transformation applied influences the amount of matrix elements to be simulated on a quantum computer and can be chosen to be only comprised of near-term-suitable expectation value measurements.

% quantum feasibility
Looking at the feasibility of the working equation's matrix elements for near-term quantum computers, together with the numerical results, revealed that the qLR formulations using projected active-space excitations with naive orbital rotations (proj LRSD) and projected active-space excitations with projected orbital rotations (all-proj LRSD) are clear candidates for future applications. Both qLR methods have a reduced amount of unique terms to be simulated for the excited state eigenvalue equation (10 and 7, respectively) compared to a naive implementation (18 terms); they only require near-term-suitable Pauli string measurements instead of expensive Hadamard test circuits; and they perform well contrasted with classic CASSCF simulations despite their active space truncation to singles and doubles.

%orbital rotation impact
Regardless of the near-term quantum feasibility, we obtained various new insights into our eight proposed qLR methods. Firstly, using different operator formalisms for the orbital rotation revealed that U-q LR methods that use self-consistent or state-transfer orbital rotation operators suffer under-parametrization and, thus, are not candidates for future quantum (or even classical) applications. The difference between naive orbital rotation operators in naive-q LR methods and projected orbital rotation operators in proj-q LR methods was more subtle. We could show that the former converges to classic CASSCF LR, while the latter is a new class of method that performed differently to classic CASSCF LR, yet not worse in comparison to classic FCI LR. 
%active-space impact
Second, the impact of the active-space truncation for all our qLR approaches was best seen for N2 \textit{(6,6)} / 6-31G, where we observed a parameter space reduction from 174 to 54 without loss of quality in the absorption spectrum. Third, as we operate with truncated active-space excitations, the transformations of the active-space excitation operators impact the results. This was seen for more challenging systems, like the highly multi-configurational stretched BeH2 \textit{(4,6)} and H4 \textit{(4,4)}. Naive LR showed a slightly worse convergence towards the complete CI active space in CASSCF than the other viable qLR methods.

Finally, we point out that for applications beyond near-term quantum computers, as well as for classic algorithms, the proposed ST-proj LR method that combines state-transfer active-space excitations with projected orbital rotations yields very compact equations with great numerical results. 

Future works will focus on simulating various other response properties, studying the effect of probabilistic and simulated noise, and focusing on a formulation of the working equations in terms of reduced density matrices. 

\section*{Appendix}

\subsection{Working equations for excited states} \label{sec:LR_WE}
Combining the various operator transformation definitions in \autoref{sec:transform} with the generalized eigenvalue equation for excited states within LR theory (\autoref{eq:LR_nopert}), gives the following working equations for the submatrices in the electronic Hessian and Metric, $\textbf{E}^{[2]}$ and $\textbf{S}^{[2]}$, respectively. Note that for all the methods, $\boldsymbol{\Delta}=\boldsymbol{0}$. As $\boldsymbol{A}$ is hermitian and $\boldsymbol{B}$ is symmetric, only one off-diagonal element is shown in the following.

\paragraph{naive LR\label{sec:naiveLR}}

The naive LR parameterization uses $\hat{G}$ and $\hat{q}$
\begin{align}
    \boldsymbol{A} &= \begin{pmatrix}
    \left<0\left|\hat{q}_\mu^\dagger\hat{H}\hat{q}_{\mu^\prime}\right|0\right> -\left<0\left|\hat{q}_\mu^\dagger\hat{q}_{\mu^\prime}\hat{H}\right|0\right> & \left(A^{Gq}\right)^\dagger \\
      \left<0\left|\hat{G}_{n}^\dagger\hat{H}\hat{q}_{\mu^\prime}\right|0\right>-\left<0\left|\hat{H}\hat{q}_{\mu^\prime}\hat{G}_{n}^\dagger\right|0\right> & \left<0\left|\left[\hat{G}_{n}^\dagger,\left[\hat{H},\hat{G}_{n^\prime}\right]\right]\right|0\right>\\        
     \end{pmatrix}\\
    \boldsymbol{B} &= \begin{pmatrix}
     -\left<0\left|\hat{q}_\mu^\dagger\hat{q}_{\mu^\prime}^{\dagger}\hat{H}\right|0\right> & \left(B^{Gq}\right)^T \\
    \left<0\left|\hat{q}_{\mu^\prime}^\dagger\hat{H}\hat{G}_{n}^\dagger\right|0\right> -\left<0\left|\hat{G}_{n}^\dagger\hat{q}_{\mu^\prime}^\dagger\hat{H}\right|0\right> & \left<0\left|\left[\hat{G}_{n}^\dagger,\left[\hat{H},\hat{G}_{n^\prime}^\dagger\right]\right]\right|0\right>\\        
     \end{pmatrix}\\
     \boldsymbol{\Sigma} &= \begin{pmatrix}
      \left<0\left|\hat{q}_\mu^\dagger\hat{q}_{\mu^\prime}\right|0\right> & 0\\
      0 & \left<0\left|\left[\hat{G}_{n}^\dagger,\hat{G}_{n^\prime}\right]\right|0\right>\\        
     \end{pmatrix}
\end{align}

\paragraph{ST LR\label{sec:stLR}} 

In ST LR, we use state-transfer active-space excitation operators, meaning $\hat{G}\rightarrow\boldsymbol{U}\hat{G}\left|\text{CSF}\right>\left<0\right|$, together with naive $q$.
This operator combination is also employed in classic MCSCF theory~\cite{olsen1985linear,Jorgensen1988} and gives
\begin{align}
    \boldsymbol{A} &= \begin{pmatrix}
     \left<0\left|\hat{q}_\mu^\dagger\hat{H}\hat{q}_{\mu^\prime}\right|0\right> -\left<0\left|\hat{q}_\mu^\dagger\hat{q}_{\mu^\prime}\hat{H}\right|0\right> & \left(A^{Gq}\right)^\dagger\\
      \left<\text{CSF}\left|\hat{G}_{n}^\dagger\boldsymbol{U}^\dagger\hat{H}\hat{q}_{\mu^\prime}\right|0\right> & \left<\text{CSF}\left|\hat{G}_{n}^\dagger\boldsymbol{U}^\dagger\hat{H}\boldsymbol{U}\hat{G}_{n^\prime}\right|\text{CSF}\right> - \delta_{nn^\prime}E_0\\        
     \end{pmatrix}\label{eq:st_A}\\
    \boldsymbol{B} &= \begin{pmatrix}
     -\left<0\left|\hat{q}_\mu^\dagger\hat{q}_{\mu^\prime}^{\dagger}\hat{H}\right|0\right> &  \left(B^{Gq}\right)^T\\
     -\left<\text{CSF}\left|\hat{G}_{n}^\dagger\boldsymbol{U}^\dagger\hat{q}_{\mu^\prime}^\dagger\hat{H}\right|0\right> & 0\\        
     \end{pmatrix}\label{eq:st_B}\\
     \boldsymbol{\Sigma} &= \begin{pmatrix}
      \left<0\left|\hat{q}_\mu^\dagger\hat{q}_{\mu^\prime}\right|0\right> & 0\\
      0 & \delta_{nn^\prime}\\        
     \end{pmatrix}
\end{align}

\paragraph{SC LR}
A transformation according to $\hat{G}\rightarrow\boldsymbol{U}\hat{G}\boldsymbol{U}^\dagger$ with naive $q$ defines SC LR
\begin{align}
    \boldsymbol{A} &= \begin{pmatrix}
     \left<0\left|\hat{q}_\mu^\dagger\hat{H}\hat{q}_{\mu^\prime}\right|0\right> -\left<0\left|\hat{q}_\mu^\dagger\hat{q}_{\mu^\prime}\hat{H}\right|0\right> & \left(A^{Gq}\right)^\dagger \\[10pt]
     \left<\text{CSF}\left|\hat{G}_{n}^\dagger\boldsymbol{U}^\dagger\hat{H}\hat{q}_{\mu^\prime}\right|0\right> & \begin{matrix}\left<\text{CSF}\left|\hat{G}_{n}^\dagger\boldsymbol{U}^\dagger\hat{H}\boldsymbol{U}\hat{G}_{n^\prime}\right|\text{CSF}\right>\\ -\left<\text{CSF}\left|G_i^\dagger\hat{G}_{n^\prime}\boldsymbol{U}^\dagger\hat{H}\right|0\right>\end{matrix}\\        
     \end{pmatrix}\label{eq:sc_A}\\
    \boldsymbol{B} &= \begin{pmatrix}
     -\left<0\left|\hat{q}_\mu^\dagger\hat{q}_{\mu^\prime}^{\dagger}\hat{H}\right|0\right> & \left(B^{Gq}\right)^T \\
     -\left<\text{CSF}\left|\hat{G}_{n}^\dagger\boldsymbol{U}^\dagger\hat{q}_{\mu^\prime}^\dagger\hat{H}\right|0\right> & \left<\text{CSF}\left|\hat{G}_{n}^\dagger\hat{G}_{n^\prime}^\dagger\boldsymbol{U}^\dagger\hat{H}\right|0\right>\\        
     \end{pmatrix}\label{eq:sc_B}\\
     \boldsymbol{\Sigma} &= \begin{pmatrix}
      \left<0\left|\hat{q}_\mu^\dagger\hat{q}_{\mu^\prime}\right|0\right> & 0\\
      0 & \delta_{nn^\prime}\\        
     \end{pmatrix}
\end{align}

\paragraph{proj LR}
For proj LR, we use $\hat{G}\rightarrow\hat{G}\left|0\right>\left<0\right| - \left<0\left|\hat{G}\right|0\right>$ and naive $q$
\begin{align}
    \boldsymbol{A} &= \begin{pmatrix}
     \left<0\left|\hat{q}_\mu^\dagger\hat{H}\hat{q}_{\mu^\prime}\right|0\right> -\left<0\left|\hat{q}_\mu^\dagger\hat{q}_{\mu^\prime}\hat{H}\right|0\right> & \left(A^{Gq}\right)^\dagger \\[10pt]
      \left<0\left|\hat{G}^\dagger_i\hat{H}\hat{q}_{\mu^\prime}\right|0\right>  - \left<0\left|\hat{G}^\dagger_i\right|0\right>\left<0\left|\hat{H}\hat{q}_{\mu^\prime}\right|0\right>  & \begin{matrix}\left<0\left|\hat{G}^\dagger_i\hat{H}\hat{G}_{n^\prime}\right|0\right> - \left<0\left|\hat{G}^\dagger_i\hat{G}_{n^\prime}\right|0\right>E_0\\ - \left<0\left|\hat{G}_{n}^\dagger\right|0\right>\left<0\left|\hat{H}\hat{G}_{n^\prime}\right|0\right>\\+\left<0\left|\hat{G}_{n}^\dagger\right|0\right>\left<0\left|\hat{G}_{n^\prime}\right|0\right>E_0\end{matrix}        
     \end{pmatrix}\\
    \boldsymbol{B} &= \begin{pmatrix}
     -\left<0\left|\hat{q}_\mu^\dagger\hat{q}_{\mu^\prime}^{\dagger}\hat{H}\right|0\right> & \left(B^{Gq}\right)^T\\
     -\left<0\left|G^\dagger_i\hat{q}_{\mu^\prime}^\dagger\hat{H}\right|0\right> + \left<0\left|\hat{q}_{\mu^\prime}^\dagger\hat{H}\right|0\right>\left<0\left|\hat{G}_{n^\prime}^\dagger\right|0\right>  & \left<0\left|\hat{G}_{n}^\dagger\hat{H}\right|0\right>\left<0\left|\hat{G}_{n^\prime}^\dagger\right|0\right> -\left<0\left|\hat{G}_{n}^\dagger\right|0\right>\left<0\left|\hat{G}_{n^\prime}^\dagger\right|0\right>E_0\\        
     \end{pmatrix}\\
     \boldsymbol{\Sigma} &= \begin{pmatrix}
      \left<0\left|\hat{q}_\mu^\dagger\hat{q}_{\mu^\prime}\right|0\right> & 0\\
      0 & \left<0\left|\hat{G}_{n}^\dagger\hat{G}_{n^\prime}\right|0\right> - \left<0\left|\hat{G}_{n}^\dagger\right|0\right>\left<0\left|\hat{G}_{n^\prime}\right|0\right>\\        
     \end{pmatrix}
\end{align}

\paragraph{all-ST LR\label{sec:all-stLR}}
The transformations $\hat{G}\rightarrow\boldsymbol{U}\hat{G}\left|\text{CSF}\right>\left<0\right|$ and $\hat{q}\rightarrow\boldsymbol{U}\hat{q}\left|\text{CSF}\right>\left<0\right|$ yield
\begin{align}
    \boldsymbol{A} &= \begin{pmatrix}
     \left<\text{CSF}\left|\hat{q}_\mu^\dagger\boldsymbol{U}^\dagger\hat{H}\boldsymbol{U}\hat{q}_{\mu^\prime}\right|\text{CSF}\right> - \delta_{\mu\mu^\prime}E_0 & \left(A^{Gq}\right)^\dagger\\
     \left<\text{CSF}\left|\hat{G}_{n}^\dagger\boldsymbol{U}^\dagger\hat{H}\boldsymbol{U}\hat{q}_{\mu^\prime}\right|\text{CSF}\right>  & \left<\text{CSF}\left|\hat{G}_{n}^\dagger\boldsymbol{U}^\dagger\hat{H}\boldsymbol{U}\hat{G}_{n^\prime}\right|\text{CSF}\right> - \delta_{nn^\prime}E_0\\        
     \end{pmatrix}\\
    \boldsymbol{B} &= \begin{pmatrix}
     0 & 0\\
      0 & 0\\        
     \end{pmatrix}\\
     \boldsymbol{\Sigma} &= \begin{pmatrix}
       \delta_{\mu\mu^\prime} & 0\\
      0 & \delta_{nn^\prime}\\        
     \end{pmatrix}
\end{align}

\paragraph{all-SC LR\label{sec:all-scLR}}
Using $\hat{G}\rightarrow\boldsymbol{U}\hat{G}\boldsymbol{U}^\dagger$ and $\hat{q}\rightarrow\boldsymbol{U}\hat{q}\boldsymbol{U}^\dagger$ yield
\begin{align}
    \boldsymbol{A} &= \begin{pmatrix}
    \begin{matrix}\left<\text{CSF}\left|\hat{q}_\mu^\dagger\boldsymbol{U}^\dagger\hat{H}\boldsymbol{U}\hat{q}_{\mu^\prime}\right|\text{CSF}\right>\\ -\left<\text{CSF}\left|\hat{q}_\mu^\dagger\hat{q}_{\mu^\prime}\boldsymbol{U}^\dagger\hat{H}\right|0\right>\end{matrix} & \left(A^{Gq}\right)^\dagger\\[10pt]
    \left<\text{CSF}\left|\hat{G}_{n}^\dagger\boldsymbol{U}^\dagger\hat{H}\boldsymbol{U}\hat{q}_{\mu^\prime}\right|\text{CSF}\right> & \begin{matrix}\left<\text{CSF}\left|\hat{G}_{n}^\dagger\boldsymbol{U}^\dagger\hat{H}\boldsymbol{U}\hat{G}_{n^\prime}\right|\text{CSF}\right>\\ -\left<\text{CSF}\left|G_i^\dagger\hat{G}_{n^\prime}\boldsymbol{U}^\dagger\hat{H}\right|0\right>\end{matrix}\\        
     \end{pmatrix}\\
    \boldsymbol{B} &= \begin{pmatrix}
     -\left<\text{CSF}\left|\hat{q}_\mu^\dagger\hat{q}_{\mu^\prime}^{\dagger}\boldsymbol{U}^\dagger\hat{H}\right|0\right> & \left(B^{Gq}\right)^T\\
     -\left<\text{CSF}\left|\hat{G}_{n}^\dagger\hat{q}_{\mu^\prime}^\dagger\boldsymbol{U}^\dagger\hat{H}\right|0\right>  & \left<\text{CSF}\left|\hat{G}_{n}^\dagger\hat{G}_{n^\prime}^\dagger\boldsymbol{U}^\dagger\hat{H}\right|0\right>\\        
     \end{pmatrix}\\
     \boldsymbol{\Sigma} &= \begin{pmatrix}
       \delta_{\mu\mu^\prime} & 0\\
      0 & \delta_{nn^\prime}\\        
     \end{pmatrix}
\end{align}

\paragraph{all-proj LR\label{sec:all-projLR}}
The $\hat{G}\rightarrow\hat{G}\left|0\right>\left<0\right| - \left<0\left|\hat{G}\right|0\right>$ and $\hat{G}\rightarrow\hat{q}\left|0\right>\left<0\right|$ parametrization results in
\begin{align}
    \boldsymbol{A} &= \begin{pmatrix}
     \left<0\left|\hat{q}_\mu^\dagger\hat{H}\hat{q}_{\mu^\prime}\right|0\right> -\left<0\left|\hat{q}_\mu^\dagger\hat{q}_{\mu^\prime}\right|0\right>E_0 & \left(A^{Gq}\right)^\dagger  \\[10pt]
      \left<0\left|\hat{G}^\dagger_i\hat{H}\hat{q}_{\mu^\prime}\right|0\right> & \begin{matrix}\left<0\left|\hat{G}^\dagger_i\hat{H}\hat{G}_{n^\prime}\right|0\right> - \left<0\left|\hat{G}^\dagger_i\hat{G}_{n^\prime}\right|0\right>E_0\\ - \left<0\left|\hat{G}_{n}^\dagger\right|0\right>\left<0\left|\hat{H}\hat{G}_{n^\prime}\right|0\right>\\+\left<0\left|\hat{G}_{n}^\dagger\right|0\right>\left<0\left|\hat{G}_{n^\prime}\right|0\right>E_0\end{matrix}        
     \end{pmatrix}\\
    \boldsymbol{B} &= \begin{pmatrix}
     0 & 0\\
      0 & \left<0\left|\hat{G}_{n}^\dagger\hat{H}\right|0\right>\left<0\left|\hat{G}_{n^\prime}^\dagger\right|0\right> -\left<0\left|\hat{G}_{n}^\dagger\right|0\right>\left<0\left|\hat{G}_{n^\prime}^\dagger\right|0\right>E_0\\        
     \end{pmatrix}\\
     \boldsymbol{\Sigma} &= \begin{pmatrix}
      \left<0\left|\hat{q}_\mu^\dagger\hat{q}_{\mu^\prime}\right|0\right> & 0\\
      0 & \left<0\left|\hat{G}_{n}^\dagger\hat{G}_{n^\prime}\right|0\right> - \left<0\left|\hat{G}_{n}^\dagger\right|0\right>\left<0\left|\hat{G}_{n^\prime}\right|0\right>\\        
     \end{pmatrix}
\end{align}

\paragraph{ST-proj LR\label{sec:st-projLR}}
In the ST-proj LR method, the active space parameters are transformed as the state-transfer operators, $\hat{G}\rightarrow\boldsymbol{U}\hat{G}\left|\text{CSF}\right>\left<0\right|$, and the orbital rotations are transformed into the projection operators, $\hat{q}\rightarrow\hat{q}\left|0\right>\left<0\right|$. 
\begin{align}
    \boldsymbol{A} &= \begin{pmatrix}
     \left<0\left|\hat{q}_\mu^\dagger\hat{H}\hat{q}_{\mu^\prime}\right|0\right> -\left<0\left|\hat{q}_\mu^\dagger\hat{q}_{\mu^\prime}\right|0\right>E_0 & \left(A^{Gq}\right)^\dagger \\
      \left<\text{CSF}\left|\hat{G}_{n}^\dagger\boldsymbol{U}^\dagger \hat{H}\hat{q}_{\mu^\prime} \right|0\right> & \left<\text{CSF}\left|\hat{G}_{n}^\dagger\boldsymbol{U}^\dagger\hat{H}\boldsymbol{U}\hat{G}_{n^\prime}\right|\text{CSF}\right> - \delta_{ij}E_0\\        
     \end{pmatrix}\label{eq:stproj_A}\\
    \boldsymbol{B} &= \begin{pmatrix}
     0 & 0\\
      0 & 0\\        
     \end{pmatrix}\label{eq:stproj_B}\\
     \boldsymbol{\Sigma} &= \begin{pmatrix}
      \left<0\left|\hat{q}_\mu^\dagger\hat{q}_{\mu^\prime}\right|0\right>  & 0\\
      0 & \delta_{nn^\prime}\\        
     \end{pmatrix}
\end{align}

\subsection{Working equations for qLR oscillator strength\label{sec:osc_WE}}
By inserting the operator transformation definitions of \autoref{sec:transform} into the expression for the oscillator strength, \autoref{eq:osc_strength}, we obtain the working equation for the state-specific property gradient, $\braket{0|[\mu_\gamma,\hat{O}_k]|0}$, and the excited state norm, $\braket{0|[\hat{O}_k,\hat{O}_k^\dagger]|0}$. 

\paragraph{naive LR}
For naive LR, we have $ \hat{X}_l \in \left\{ \hat{q}_\mu, \hat{G}_n \right\}$ for the excitation operator in \autoref{eq:exc_op} giving
\begin{align}
\nonumber
    \left<k|k\right> &= \left<0\left| \left[ \hat{O}_k, \hat{O}_k^\dagger \right] \right|0\right> \\
    &= \sum_{\substack{l \in \mu, n\\l^\prime \in \mu^\prime, n^\prime}} \Big\{ Z_{k,\mu} Z_{k,\mu^\prime}^* \left<0\left| \hat{q}_\mu^\dagger \hat{q}_{\mu^\prime} \right|0\right> - Y_{k,\mu} Y_{k,\mu^\prime}^* \left<0\left| \hat{q}_{\mu^\prime}^\dagger \hat{q}_{\mu} \right|0\right> \\\nonumber
    & \phantom{ \sum_{\substack{l \in \mu, n\\l^\prime \in \mu^\prime, n^\prime}} \Big\{} + Z_{k,n} Z_{k,n^\prime}^* \left<0\left| \hat{G}_n^\dagger \hat{G}_{n^\prime} \right|0\right>  - Z_{k,n} Z_{k,n^\prime}^* \left<0\left| \hat{G}_{n^\prime} \hat{G}_n^\dagger \right|0\right> \\\nonumber
    & \phantom{ \sum_{\substack{l \in \mu, n\\l^\prime \in \mu^\prime, n^\prime}} \Big\{} + Y_{k,n} Y_{k,n^\prime}^* \left<0\left| \hat{G}_n \hat{G}_{n^\prime}^\dagger \right|0\right>  - Y_{k,n} Y_{k,n^\prime}^* \left<0\left| \hat{G}_{n^\prime}^\dagger \hat{G}_n \right|0\right> \Big\} \\
    \left<0\left|\left[\hat{\mu}_\gamma,\hat{O}_k\right]\right|0\right> &= \sum_{l \in \mu, n} \Big\{ - Z_{k,\mu} \left<0\left| \hat{q}_\mu^\dagger \hat{\mu}_\gamma \right|0\right>  + Z_{k,n} \left<0\left| \hat{\mu}_\gamma \hat{G}_n^\dagger  \right|0\right> - Z_{k,n} \left<0\left|  \hat{G}_n^\dagger  \hat{\mu}_\gamma \right|0\right> \\\nonumber
     &\phantom{ \sum_{l \in \mu, n} \Big\{} + Y_{k,\mu} \left<0\left|  \hat{\mu}_\gamma \hat{q}_\mu \right|0\right>  + Y_{k,n} \left<0\left| \hat{\mu}_\gamma \hat{G}_n  \right|0\right> - Y_{k,n} \left<0\left|  \hat{G}_n  \hat{\mu}_\gamma \right|0\right>\Big\}
\end{align}

\paragraph{proj LR and all-proj LR}
For proj LR and all-proj LR, we have $\hat{X}_l \in \left\{ \hat{q}_\mu, \hat{G}_n \ket{0}\bra{0} -\braket{0|\hat{G}_n|0} \right\}$ and $\hat{X}_l \in \left\{ \hat{q}_\mu \ket{0}\bra{0}, \hat{G}_n \ket{0}\bra{0} -\braket{0|\hat{G}_n|0}\right\}$, respectively. They give the same working equations,    
\begin{align}
\nonumber
    \left<k|k\right> &= \left<0\left| \left[ \hat{O}_k, \hat{O}_k^\dagger \right] \right|0\right> \\
    &= \sum_{\substack{l \in \mu, n\\l^\prime \in \mu^\prime, n^\prime}} \Big\{ Z_{k,\mu} Z_{k,\mu^\prime}^* \left<0\left| \hat{q}_\mu^\dagger \hat{q}_{\mu^\prime} \right|0\right> - Y_{k,\mu} Y_{k,\mu^\prime}^* \left<0\left| \hat{q}_{\mu^\prime}^\dagger \hat{q}_{\mu} \right|0\right> \\\nonumber
    & \phantom{ \sum_{\substack{l \in \mu, n\\l^\prime \in \mu^\prime, n^\prime}} \Big\{} + Z_{k,n} Z_{k,n^\prime}^* \left<0\left| \hat{G}_n^\dagger \hat{G}_{n^\prime} \right|0\right>  - Z_{k,n} Z_{k,n^\prime}^* \left<0\left| \hat{G}_{n^\prime} \right|0\right> \left<0\left| \hat{G}_n^\dagger \right|0\right> \\\nonumber
    & \phantom{ \sum_{\substack{l \in \mu, n\\l^\prime \in \mu^\prime, n^\prime}} \Big\{} + Y_{k,n} Y_{k,n^\prime}^* \left<0\left| \hat{G}_n \right|0\right>\left<0\left| \hat{G}_{n^\prime}^\dagger \right|0\right>  - Y_{k,n} Y_{k,n^\prime}^* \left<0\left| \hat{G}_{n^\prime}^\dagger \hat{G}_n \right|0\right> \Big\} \\
    \nonumber
    \left<0\left|\left[\hat{\mu}_\gamma,\hat{O}_k\right]\right|0\right> &= \sum_{l \in \mu, n} \Big\{ - Z_{k,\mu} \left<0\left| \hat{q}_\mu^\dagger \hat{\mu}_\gamma \right|0\right>  + Z_{k,n} \left<0\left| \hat{\mu}_\gamma \right|0\right> \left<0\left|  \hat{G}_n^\dagger  \right|0\right> - Z_{k,n} \left<0\left|  \hat{G}_n^\dagger  \hat{\mu}_\gamma \right|0\right> \\
     &\phantom{ \sum_{l \in \mu, n} \Big\{} + Y_{k,\mu} \left<0\left|  \hat{\mu}_\gamma \hat{q}_\mu \right|0\right>  + Y_{k,n} \left<0\left| \hat{\mu}_\gamma \hat{G}_n  \right|0\right> - Y_{k,n} \left<0\left|  \hat{G}_n \right|0\right> \left<0\left| \hat{\mu}_\gamma \right|0\right>\Big\}
\end{align}

\paragraph{ST-LR, SC-LR, ST-proj-LR}
For ST LR, SC LR, and ST-proj LR, we have $\hat{X}_l \in \left\{ \hat{q}_\mu, \boldsymbol{U}\hat{G}_n\left|\text{CSF}\right>\left<0\right| \right\}$, $\hat{X}_l \in \left\{ \hat{q}_\mu, \boldsymbol{U}\hat{G}_n\boldsymbol{U}^\dagger \right\}$, and $ \hat{X} \in \left\{ \hat{q} \ket{0}\bra{0} , \boldsymbol{U}\hat{G}\left|\text{CSF}\right>\left<0\right| \right\}$, respectively. They give the same working equations,    
\begin{align}
\nonumber
    \left<k|k\right> &= \left<0\left| \left[ \hat{O}_k, \hat{O}_k^\dagger \right] \right|0\right> \\\nonumber
    &= \sum_{\substack{l \in \mu, n\\l^\prime \in \mu^\prime, n^\prime}} \Big\{ Z_{k,\mu} Z_{k,\mu^\prime}^* \left<0\left| \hat{q}_\mu^\dagger \hat{q}_{\mu^\prime} \right|0\right> - Y_{k,\mu} Y_{k,\mu^\prime}^* \left<0\left| \hat{q}_{\mu^\prime}^\dagger \hat{q}_{\mu} \right|0\right> \\
    & \phantom{ \sum_{\substack{l \in \mu, n\\l^\prime \in \mu^\prime, n^\prime}} \Big\{} + \left|Z_{k,n}\right|^2 - \left|Y_{k,n}\right|^2 \\
    \nonumber
    \left<0\left|\left[\hat{\mu}_\gamma,\hat{O}_k\right]\right|0\right> &= \sum_{l \in \mu, n} \Big\{ - Z_{k,\mu} \left<0\left| \hat{q}_\mu^\dagger \hat{\mu}_\gamma \right|0\right> - Z_{k,n} \left<\text{CSF}\left|  \hat{G}_n^\dagger \boldsymbol{U}^\dagger \hat{\mu}_\gamma \right|0\right> \\
     &\phantom{ \sum_{l \in \mu, n} \Big\{} + Y_{k,\mu} \left<0\left|  \hat{\mu}_\gamma \hat{q}_\mu \right|0\right>  + Y_{k,n} \left<0\left| \hat{\mu}_\gamma \boldsymbol{U} \hat{G}_n  \right|\text{CSF}\right> \Big\}
\end{align}

\paragraph{all-ST LR, all-SC LR} 
For all-ST LR and all-SC LR, we have $\hat{X}_l \in \left\{ \boldsymbol{U}\hat{q}_\mu \left|\text{CSF}\right>\left<0\right|, \boldsymbol{U}\hat{G}_n\left|\text{CSF}\right>\left<0\right| \right\}$, and $\hat{X}_l \in \left\{ \boldsymbol{U}\hat{q}_\mu \boldsymbol{U}^\dagger, \boldsymbol{U}\hat{G}_n\boldsymbol{U}^\dagger \right\}$, respectively. They give the same working equations,    
\begin{align}
\nonumber
    \left<k|k\right> &= \left<0\left| \left[ \hat{O}_k, \hat{O}_k^\dagger \right] \right|0\right> \\
    &= \sum_{l \in \mu, n} \Big\{ \left|Z_{k,\mu}\right|^2  - \left|Y_{k,\mu}\right|^2 + \left|Z_{k,n}\right|^2 - \left|Y_{k,n}\right|^2 \Big\}\\
    \nonumber
    \left<0\left|\left[\hat{\mu}_\gamma,\hat{O}_k\right]\right|0\right> &= \sum_{l \in \mu, n} \Big\{ - Z_{k,\mu} \left<\text{CSF}\left| \hat{q}_\mu^\dagger \boldsymbol{U}^\dagger \hat{\mu}_\gamma \right|0\right> - Z_{k,n} \left<\text{CSF}\left|  \hat{G}_n^\dagger \boldsymbol{U}^\dagger \hat{\mu}_\gamma \right|0\right> \\
     &\phantom{ \sum_{l \in \mu, n} \Big\{} + Y_{k,\mu} \left<0\left|  \hat{\mu}_\gamma \boldsymbol{U} \hat{q}_\mu \right|\text{CSF}\right>  + Y_{k,n} \left<0\left| \hat{\mu}_\gamma \boldsymbol{U} \hat{G}_n  \right|\text{CSF}\right> \Big\}
\end{align}

\section{Additional figures}

%\subsection{H$_4$ \textit{(4,4)}}
\begin{figure}[htbp!]
	\centering 
	\includegraphics[width=.7\textwidth]{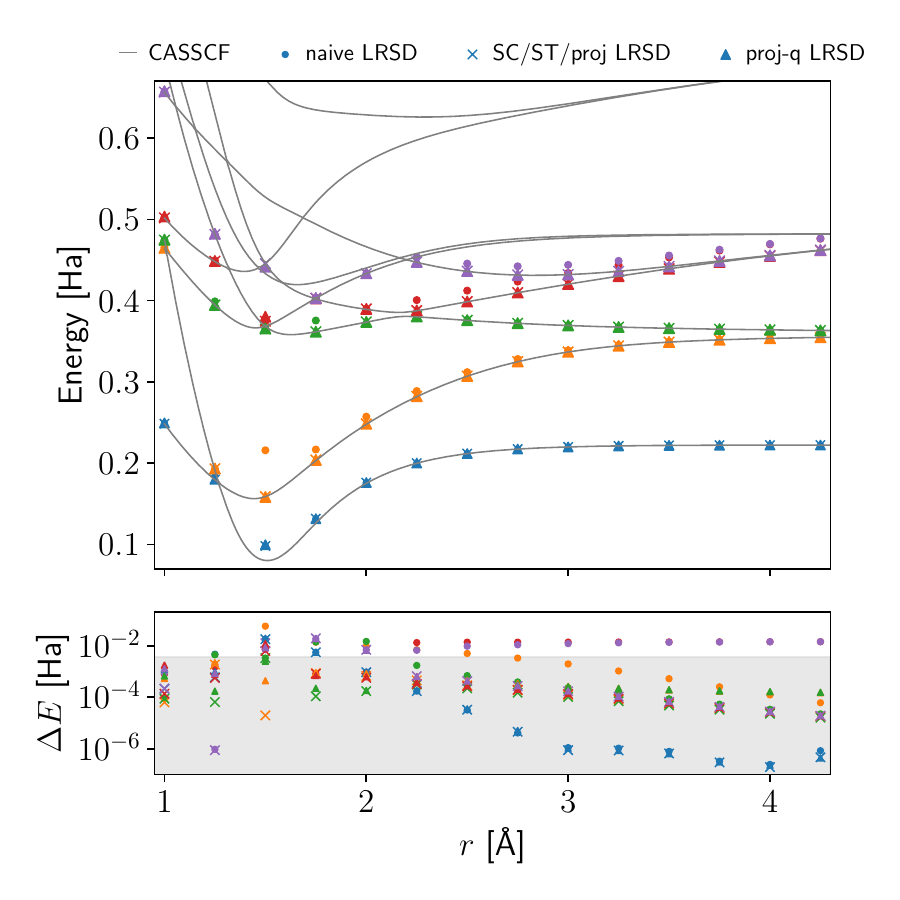}
	\caption{Excited states of H$_4$ \textit{(4,4)} / cc-pVDZ. Comparison of naive LRSD (dots) SC/ST/proj LRSD (crosses), and proj-q LRSD (triangles) with CASSCF (gray line).}
	\label{fig:H4_all}
\end{figure}

%\subsection{H$_2$O \textit{(4,4)}}
\begin{figure}[htbp!]
	\centering 
        \subfigure{
	\includegraphics[width=.5\textwidth]{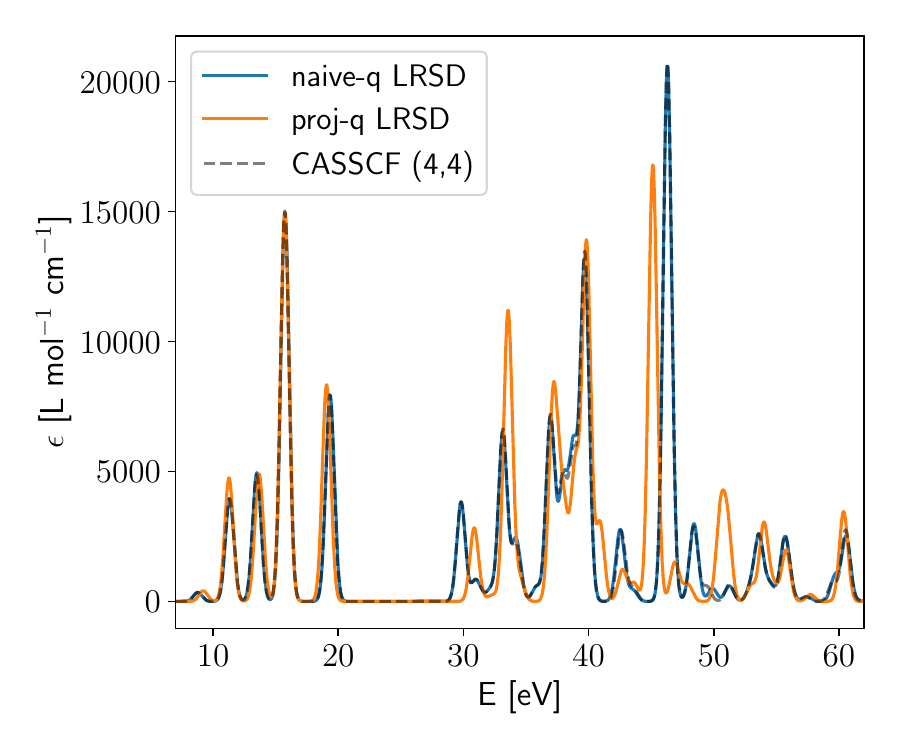}}%
        \subfigure{
	\includegraphics[width=.5\textwidth]{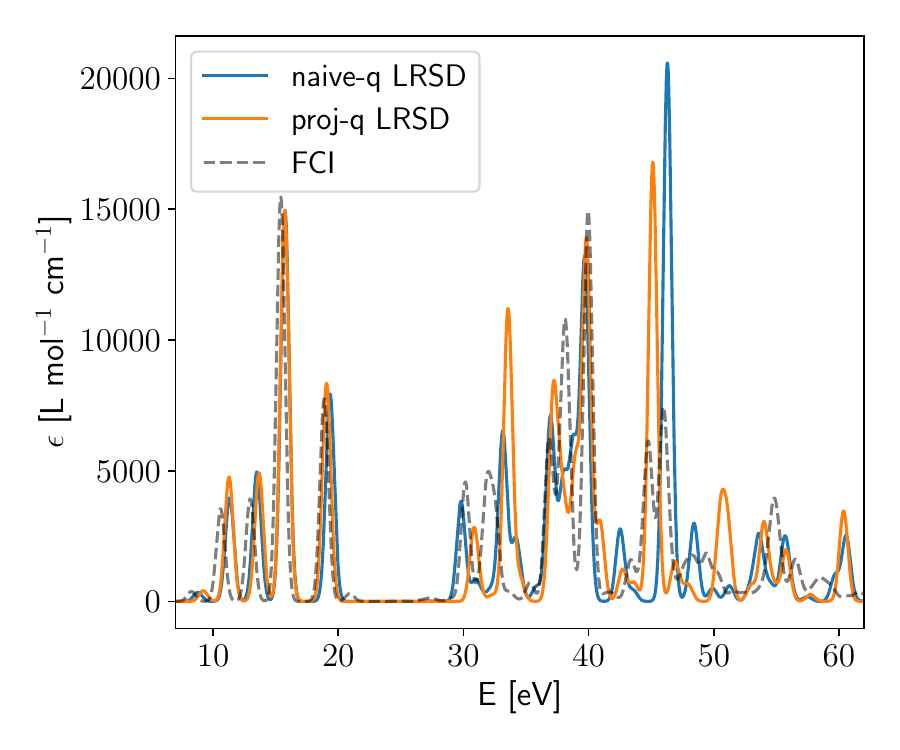}}
	\caption{Absorption spectra of H$_2$O \textit{(4,4)} / 6-31G with various qLR methods compared to CASSCF \textit{(4,4)} (left) and FCI (right).}
	\label{fig:H2O_eq_631G_fullrange}
\end{figure}

%\subsection{BeH$_2$ \textit{(4,6)}}
\begin{figure}[htbp!]
	\centering 
	\subfigure{
    \includegraphics[width=.5\textwidth]{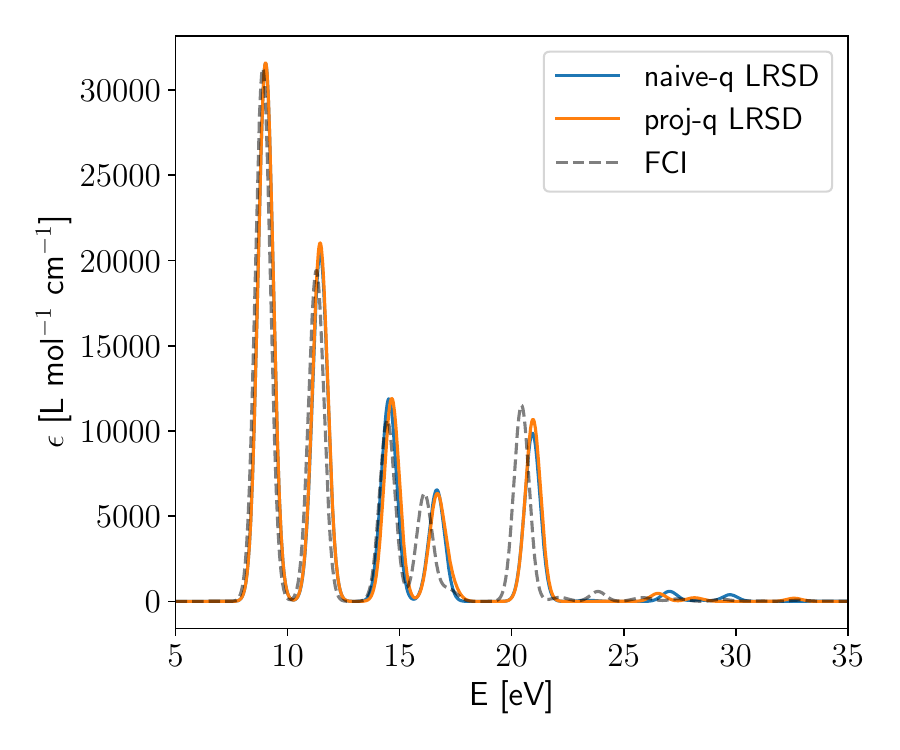}}%
    \subfigure{
	\includegraphics[width=.5\textwidth]{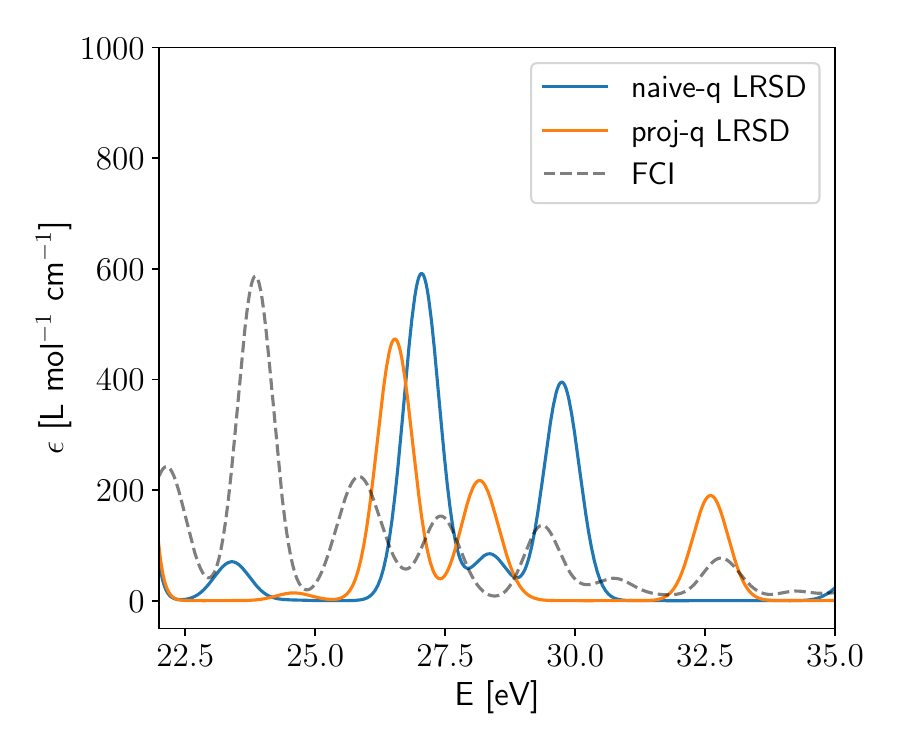}}
	\caption{Absorption spectra of BeH$_2$ \textit{(4,6)} / 6-31G using various qLR methods compared to FCI in full energy range (left) and in selected range (right).}
	\label{fig:BeH2_eq_46_631G_FCI}
\end{figure}

\begin{figure}[htbp!]
	\centering 
	\includegraphics[width=0.7\textwidth]{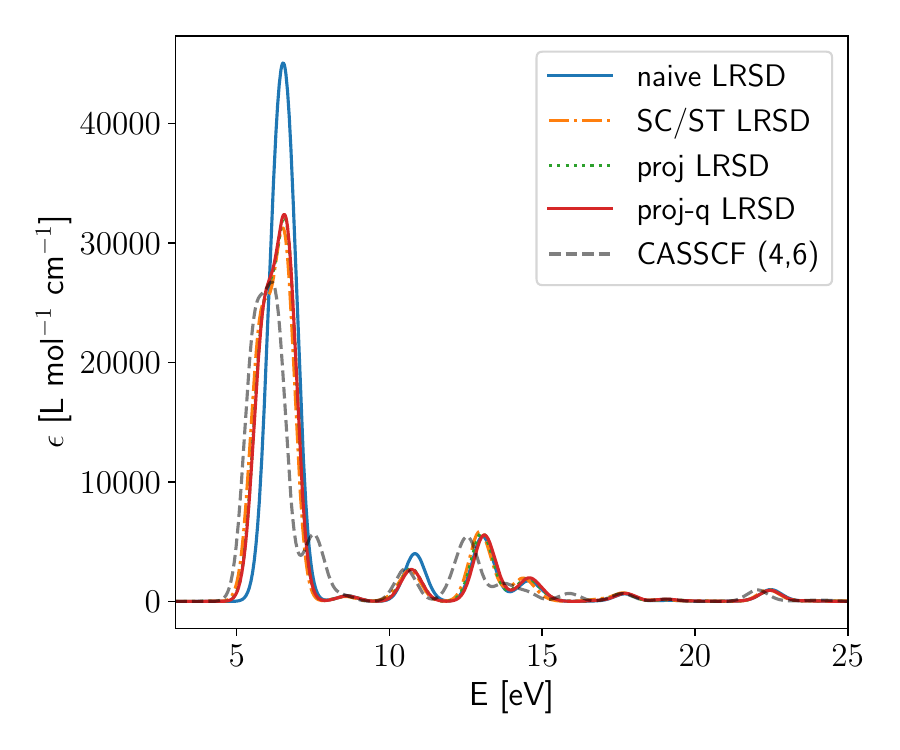}%
	\caption{Symmetrically stretched (doubled bond length) BeH$_2$ \textit{(4,6)} / 6-31G, various qLR methods compared to CASSCF\textit{(4,6)} in full range (left) and in the selected range (right).}
	\label{fig:BeH2_str_46_631G_CAS_full}
\end{figure}

\begin{figure}[htbp!]
	\centering 
        \subfigure{
	\includegraphics[width=.5\textwidth]{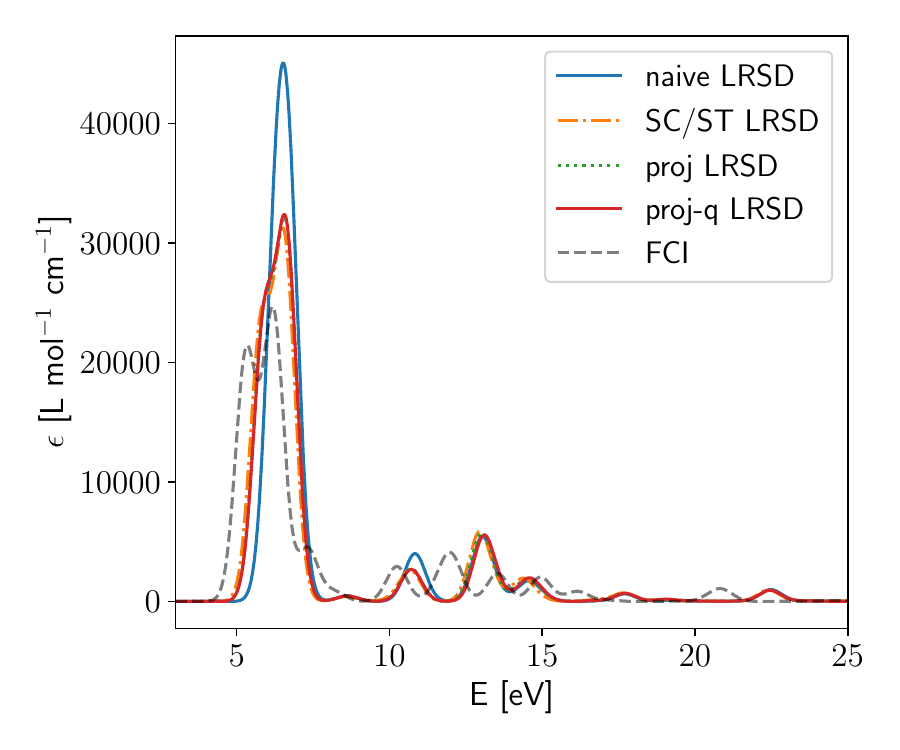}}%
        \subfigure{
	\includegraphics[width=.5\textwidth]{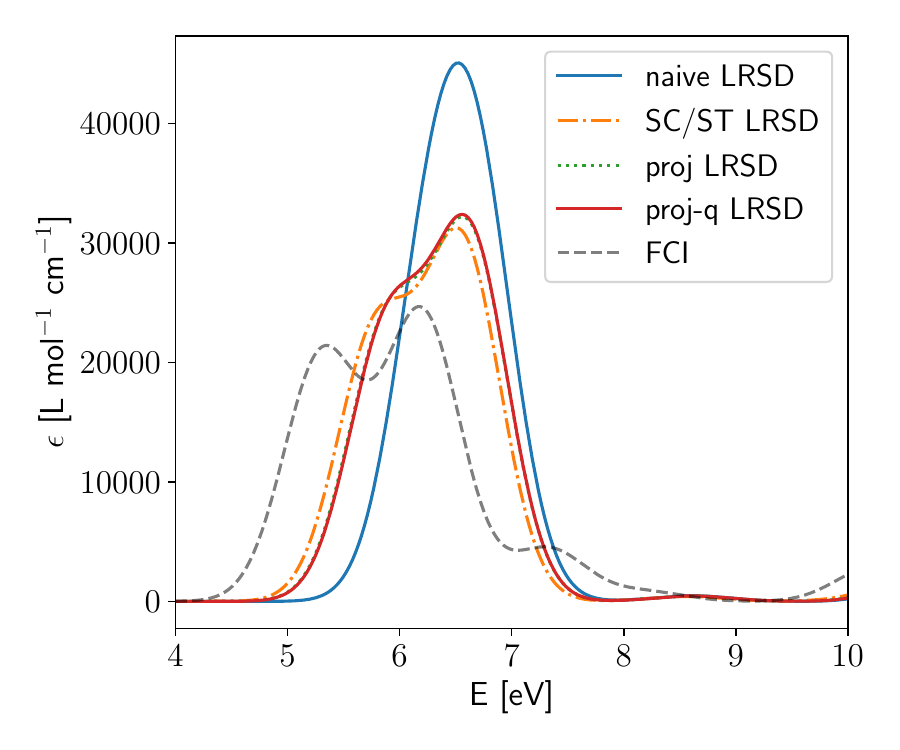}}
	\caption{Symmetrically stretched BeH$_2$ (4,6) / 6-31G, various qLR methods compared to FCI in full energy range (left) and in selected range(right).}
	\label{fig:BeH2_str_46_631G_FCI}
\end{figure}

\section*{Acknowledgments}
The authors are grateful for fruitful discussions with Theo Juncker von Buchwald (DTU) and Stefan Knecht (Algorithmiq).
Financial Support from the Novo Nordisk Foundation (NNF) for the focused research project ``Hybrid Quantum Chemistry on Hybrid Quantum Computers'' (NNF grant NNFSA220080996) is acknowledged.

%\section*{DATA AVAILABILITY STATEMENT}

%TBD

\bibliography{literature}

\providecommand{\latin}[1]{#1}
\makeatletter
\providecommand{\doi}
  {\begingroup\let\do\@makeother\dospecials
  \catcode`\{=1 \catcode`\}=2 \doi@aux}
\providecommand{\doi@aux}[1]{\endgroup\texttt{#1}}
\makeatother
\providecommand*\mcitethebibliography{\thebibliography}
\csname @ifundefined\endcsname{endmcitethebibliography}  {\let\endmcitethebibliography\endthebibliography}{}
\begin{mcitethebibliography}{62}
\providecommand*\natexlab[1]{#1}
\providecommand*\mciteSetBstSublistMode[1]{}
\providecommand*\mciteSetBstMaxWidthForm[2]{}
\providecommand*\mciteBstWouldAddEndPuncttrue
  {\def\EndOfBibitem{\unskip.}}
\providecommand*\mciteBstWouldAddEndPunctfalse
  {\let\EndOfBibitem\relax}
\providecommand*\mciteSetBstMidEndSepPunct[3]{}
\providecommand*\mciteSetBstSublistLabelBeginEnd[3]{}
\providecommand*\EndOfBibitem{}
\mciteSetBstSublistMode{f}
\mciteSetBstMaxWidthForm{subitem}{(\alph{mcitesubitemcount})}
\mciteSetBstSublistLabelBeginEnd
  {\mcitemaxwidthsubitemform\space}
  {\relax}
  {\relax}

\bibitem[Olsen and J{\o}rgensen(1985)Olsen, and J{\o}rgensen]{olsen1985linear}
Olsen,~J.; J{\o}rgensen,~P. Linear and nonlinear response functions for an exact state and for an MCSCF state. \emph{The Journal of Chemical Physics} \textbf{1985}, \emph{82}, 3235--3264\relax
\mciteBstWouldAddEndPuncttrue
\mciteSetBstMidEndSepPunct{\mcitedefaultmidpunct}
{\mcitedefaultendpunct}{\mcitedefaultseppunct}\relax
\EndOfBibitem
\bibitem[Christiansen \latin{et~al.}(1998)Christiansen, J{\o}rgensen, and H{\"a}ttig]{Christiansen1998}
Christiansen,~O.; J{\o}rgensen,~P.; H{\"a}ttig,~C. Response functions from Fourier component variational perturbation theory applied to a time-averaged quasienergy. \emph{International Journal of Quantum Chemistry} \textbf{1998}, \emph{68}, 1--52\relax
\mciteBstWouldAddEndPuncttrue
\mciteSetBstMidEndSepPunct{\mcitedefaultmidpunct}
{\mcitedefaultendpunct}{\mcitedefaultseppunct}\relax
\EndOfBibitem
\bibitem[Helgaker \latin{et~al.}(2012)Helgaker, Coriani, J{\o}rgensen, Kristensen, Olsen, and Ruud]{helgaker2012recent}
Helgaker,~T.; Coriani,~S.; J{\o}rgensen,~P.; Kristensen,~K.; Olsen,~J.; Ruud,~K. Recent advances in wave function-based methods of molecular-property calculations. \emph{Chemical Reviews} \textbf{2012}, \emph{112}, 543--631\relax
\mciteBstWouldAddEndPuncttrue
\mciteSetBstMidEndSepPunct{\mcitedefaultmidpunct}
{\mcitedefaultendpunct}{\mcitedefaultseppunct}\relax
\EndOfBibitem
\bibitem[Paw{\l}owski \latin{et~al.}(2015)Paw{\l}owski, Olsen, and J{\o}rgensen]{Pawlowski2015}
Paw{\l}owski,~F.; Olsen,~J.; J{\o}rgensen,~P. Molecular response properties from a {Hermitian} eigenvalue equation for a time-periodic {Hamiltonian}. \emph{The Journal of Chemical Physic} \textbf{2015}, \emph{142}, 114109\relax
\mciteBstWouldAddEndPuncttrue
\mciteSetBstMidEndSepPunct{\mcitedefaultmidpunct}
{\mcitedefaultendpunct}{\mcitedefaultseppunct}\relax
\EndOfBibitem
\bibitem[Aidas \latin{et~al.}(2013)Aidas, Angeli, Bak, Bakken, Bast, Boman, Christiansen, Cimiraglia, Coriani, Dahle, Dalskov, Ekstr\"om, Enevoldsen, Eriksen, Ettenhuber, Fern\'andez, Ferrighi, Fliegl, Frediani, Hald, Halkier, H\"attig, Heiberg, Helgaker, Hennum, Hettema, Hjertenaes, H{\o}st, H{\o}yvik, Iozzi, Jans{\'\i}k, Jensen, Jonsson, J{\o}rgensen, Kauczor, Kirpekar, Kjaergaard, Klopper, Knecht, Kobayashi, Koch, Kongsted, Krapp, Kristensen, Ligabue, Lutnaes, Melo, Mikkelsen, Myhre, Neiss, Nielsen, Norman, Olsen, Olsen, Osted, Packer, Pawlowski, Pedersen, Provasi, Reine, Rinkevicius, Ruden, Ruud, Rybkin, Sa{\l}ek, Samson, de~Mer\'as, Saue, Sauer, Schimmelpfennig, Sneskov, Steindal, Sylvester-Hvid, Taylor, Teale, Tellgren, Tew, Thorvaldsen, Th{\o}gersen, Vahtras, Watson, Wilson, Ziolkowski, and \r{A}gren]{Aidas2013}
Aidas,~K.; Angeli,~C.; Bak,~K.~L.; Bakken,~V.; Bast,~R.; Boman,~L.; Christiansen,~O.; Cimiraglia,~R.; Coriani,~S.; Dahle,~P.; Dalskov,~E.~K.; Ekstr\"om,~U.; Enevoldsen,~T.; Eriksen,~J.~J.; Ettenhuber,~P.; Fern\'andez,~B.; Ferrighi,~L.; Fliegl,~H.; Frediani,~L.; Hald,~K.; Halkier,~A.; H\"attig,~C.; Heiberg,~H.; Helgaker,~T.; Hennum,~A.~C.; Hettema,~H.; Hjertenaes,~E.; H{\o}st,~S.; H{\o}yvik,~I.-M.; Iozzi,~M.~F.; Jans{\'\i}k,~B.; Jensen,~H. J.~A.; Jonsson,~D.; J{\o}rgensen,~P.; Kauczor,~J.; Kirpekar,~S.; Kjaergaard,~T.; Klopper,~W.; Knecht,~S.; Kobayashi,~R.; Koch,~H.; Kongsted,~J.; Krapp,~A.; Kristensen,~K.; Ligabue,~A.; Lutnaes,~O.~B.; Melo,~J.~I.; Mikkelsen,~K.~V.; Myhre,~R.~H.; Neiss,~C.; Nielsen,~C.~B.; Norman,~P.; Olsen,~J.; Olsen,~J. M.~H.; Osted,~A.; Packer,~M.~J.; Pawlowski,~F.; Pedersen,~T.~B.; Provasi,~P.~F.; Reine,~S.; Rinkevicius,~Z.; Ruden,~T.~A.; Ruud,~K.; Rybkin,~V.~V.; Sa{\l}ek,~P.; Samson,~C. C.~M.; de~Mer\'as,~A.~S.; Saue,~T.; Sauer,~S. P.~A.; Schimmelpfennig,~B.; Sneskov,~K.;
  Steindal,~A.~H.; Sylvester-Hvid,~K.~O.; Taylor,~P.~R.; Teale,~A.~M.; Tellgren,~E.~I.; Tew,~D.~P.; Thorvaldsen,~A.~J.; Th{\o}gersen,~L.; Vahtras,~O.; Watson,~M.~A.; Wilson,~D. J.~D.; Ziolkowski,~M.; \r{A}gren,~H. The {Dalton} quantum chemistry program system. \emph{WIREs Comput Mol Sci} \textbf{2013}, \emph{4}, 269--284\relax
\mciteBstWouldAddEndPuncttrue
\mciteSetBstMidEndSepPunct{\mcitedefaultmidpunct}
{\mcitedefaultendpunct}{\mcitedefaultseppunct}\relax
\EndOfBibitem
\bibitem[DAL(2022)]{DALTON2022}
{DALTON, a molecular electronic structure program, Release Dalton2022, see \url{http://daltonprogram.org} }. 2022\relax
\mciteBstWouldAddEndPuncttrue
\mciteSetBstMidEndSepPunct{\mcitedefaultmidpunct}
{\mcitedefaultendpunct}{\mcitedefaultseppunct}\relax
\EndOfBibitem
\bibitem[Epifanovsky \latin{et~al.}(2021)Epifanovsky, Gilbert, Feng, Lee, Mao, Mardirossian, Pokhilko, White, Coons, Dempwolff, Gan, Hait, Horn, Jacobson, Kaliman, Kussmann, Lange, Lao, Levine, Liu, McKenzie, Morrison, Nanda, Plasser, Rehn, Vidal, You, Zhu, Alam, Albrecht, Aldossary, Alguire, Andersen, Athavale, Barton, Begam, Behn, Bellonzi, Bernard, Berquist, Burton, Carreras, Carter-Fenk, Chakraborty, Chien, Closser, Cofer-Shabica, Dasgupta, de~Wergifosse, Deng, Diedenhofen, Do, Ehlert, Fang, Fatehi, Feng, Friedhoff, Gayvert, Ge, Gidofalvi, Goldey, Gomes, Gonz{\'a}lez-Espinoza, Gulania, Gunina, Hanson-Heine, Harbach, Hauser, Herbst, Hernández~Vera, Hodecker, Holden, Houck, Huang, Hui, Huynh, Ivanov, J\'asz, Ji, Jiang, Kaduk, K{\"a}hler, Khistyaev, Kim, Kis, Klunzinger, Koczor-Benda, Koh, Kosenkov, Koulias, Kowalczyk, Krauter, Kue, Kunitsa, Kus, Ladj{\'a}nszki, Landau, Lawler, Lefrancois, Lehtola, Li, Li, Liang, Liebenthal, Lin, Lin, Liu, Liu, Loipersberger, Luenser, Manjanath, Manohar, Mansoor, Manzer,
  Mao, Marenich, Markovich, Mason, Maurer, McLaughlin, Menger, Mewes, Mewes, Morgante, Mullinax, Oosterbaan, Paran, Paul, Paul, Pavo{\v s}evi{\'c}, Pei, Prager, Proynov, R{\'a}k, Ramos-Cordoba, Rana, Rask, Rettig, Richard, Rob, Rossomme, Scheele, Scheurer, Schneider, Sergueev, Sharada, Skomorowski, Small, Stein, Su, Sundstrom, Tao, Thirman, Tornai, Tsuchimochi, Tubman, Veccham, Vydrov, Wenzel, Witte, Yamada, Yao, Yeganeh, Yost, Zech, Zhang, Zhang, Zhang, Zuev, Aspuru-Guzik, Bell, Besley, Bravaya, Brooks, Casanova, Chai, Coriani, Cramer, Cserey, DePrince, DiStasio, Dreuw, Dunietz, Furlani, Goddard, Hammes-Schiffer, Head-Gordon, Hehre, Hsu, Jagau, Jung, Klamt, Kong, Lambrecht, Liang, Mayhall, McCurdy, Neaton, Ochsenfeld, Parkhill, Peverati, Rassolov, Shao, Slipchenko, Stauch, Steele, Subotnik, Thom, Tkatchenko, Truhlar, Van~Voorhis, Wesolowski, Whaley, Woodcock, Zimmerman, Faraji, Gill, Head-Gordon, Herbert, and Krylov]{Qchem541}
Epifanovsky,~E.; Gilbert,~A. T.~B.; Feng,~X.; Lee,~J.; Mao,~Y.; Mardirossian,~N.; Pokhilko,~P.; White,~A.~F.; Coons,~M.~P.; Dempwolff,~A.~L.; Gan,~Z.; Hait,~D.; Horn,~P.~R.; Jacobson,~L.~D.; Kaliman,~I.; Kussmann,~J.; Lange,~A.~W.; Lao,~K.~U.; Levine,~D.~S.; Liu,~J.; McKenzie,~S.~C.; Morrison,~A.~F.; Nanda,~K.~D.; Plasser,~F.; Rehn,~D.~R.; Vidal,~M.~L.; You,~Z.-Q.; Zhu,~Y.; Alam,~B.; Albrecht,~B.~J.; Aldossary,~A.; Alguire,~E.; Andersen,~J.~H.; Athavale,~V.; Barton,~D.; Begam,~K.; Behn,~A.; Bellonzi,~N.; Bernard,~Y.~A.; Berquist,~E.~J.; Burton,~H. G.~A.; Carreras,~A.; Carter-Fenk,~K.; Chakraborty,~R.; Chien,~A.~D.; Closser,~K.~D.; Cofer-Shabica,~V.; Dasgupta,~S.; de~Wergifosse,~M.; Deng,~J.; Diedenhofen,~M.; Do,~H.; Ehlert,~S.; Fang,~P.-T.; Fatehi,~S.; Feng,~Q.; Friedhoff,~T.; Gayvert,~J.; Ge,~Q.; Gidofalvi,~G.; Goldey,~M.; Gomes,~J.; Gonz{\'a}lez-Espinoza,~C.~E.; Gulania,~S.; Gunina,~A.~O.; Hanson-Heine,~M. W.~D.; Harbach,~P. H.~P.; Hauser,~A.; Herbst,~M.~F.; Hernández~Vera,~M.; Hodecker,~M.; Holden,~Z.~C.;
  Houck,~S.; Huang,~X.; Hui,~K.; Huynh,~B.~C.; Ivanov,~M.; J\'asz,~A.; Ji,~H.; Jiang,~H.; Kaduk,~B.; K{\"a}hler,~S.; Khistyaev,~K.; Kim,~J.; Kis,~G.; Klunzinger,~P.; Koczor-Benda,~Z.; Koh,~J.~H.; Kosenkov,~D.; Koulias,~L.; Kowalczyk,~T.; Krauter,~C.~M.; Kue,~K.; Kunitsa,~A.; Kus,~T.; Ladj{\'a}nszki,~I.; Landau,~A.; Lawler,~K.~V.; Lefrancois,~D.; Lehtola,~S.; Li,~R.~R.; Li,~Y.-P.; Liang,~J.; Liebenthal,~M.; Lin,~H.-H.; Lin,~Y.-S.; Liu,~F.; Liu,~K.-Y.; Loipersberger,~M.; Luenser,~A.; Manjanath,~A.; Manohar,~P.; Mansoor,~E.; Manzer,~S.~F.; Mao,~S.-P.; Marenich,~A.~V.; Markovich,~T.; Mason,~S.; Maurer,~S.~A.; McLaughlin,~P.~F.; Menger,~M. F. S.~J.; Mewes,~J.-M.; Mewes,~S.~A.; Morgante,~P.; Mullinax,~J.~W.; Oosterbaan,~K.~J.; Paran,~G.; Paul,~A.~C.; Paul,~S.~K.; Pavo{\v s}evi{\'c},~F.; Pei,~Z.; Prager,~S.; Proynov,~E.~I.; R{\'a}k,~{\'A}.; Ramos-Cordoba,~E.; Rana,~B.; Rask,~A.~E.; Rettig,~A.; Richard,~R.~M.; Rob,~F.; Rossomme,~E.; Scheele,~T.; Scheurer,~M.; Schneider,~M.; Sergueev,~N.; Sharada,~S.~M.;
  Skomorowski,~W.; Small,~D.~W.; Stein,~C.~J.; Su,~Y.-C. \latin{et~al.}  Software for the frontiers of quantum chemistry: An overview of developments in the Q-Chem 5 package. \emph{The Journal of Chemical Physics} \textbf{2021}, \emph{155}, 084801\relax
\mciteBstWouldAddEndPuncttrue
\mciteSetBstMidEndSepPunct{\mcitedefaultmidpunct}
{\mcitedefaultendpunct}{\mcitedefaultseppunct}\relax
\EndOfBibitem
\bibitem[Franzke \latin{et~al.}(2023)Franzke, Holzer, Andersen, Begušić, Bruder, Coriani, Della~Sala, Fabiano, Fedotov, Fürst, Gillhuber, Grotjahn, Kaupp, Kehry, Krstić, Mack, Majumdar, Nguyen, Parker, Pauly, Pausch, Perlt, Phun, Rajabi, Rappoport, Samal, Schrader, Sharma, Tapavicza, Treß, Voora, Wodyński, Yu, Zerulla, Furche, Hättig, Sierka, Tew, and Weigend]{Turbomole2023}
Franzke,~Y.~J.; Holzer,~C.; Andersen,~J.~H.; Begušić,~T.; Bruder,~F.; Coriani,~S.; Della~Sala,~F.; Fabiano,~E.; Fedotov,~D.~A.; Fürst,~S.; Gillhuber,~S.; Grotjahn,~R.; Kaupp,~M.; Kehry,~M.; Krstić,~M.; Mack,~F.; Majumdar,~S.; Nguyen,~B.~D.; Parker,~S.~M.; Pauly,~F.; Pausch,~A.; Perlt,~E.; Phun,~G.~S.; Rajabi,~A.; Rappoport,~D.; Samal,~B.; Schrader,~T.; Sharma,~M.; Tapavicza,~E.; Treß,~R.~S.; Voora,~V.; Wodyński,~A.; Yu,~J.~M.; Zerulla,~B.; Furche,~F.; Hättig,~C.; Sierka,~M.; Tew,~D.~P.; Weigend,~F. TURBOMOLE: Today and Tomorrow. \emph{Journal of Chemical Theory and Computation} \textbf{2023}, \emph{19}, 6859--6890\relax
\mciteBstWouldAddEndPuncttrue
\mciteSetBstMidEndSepPunct{\mcitedefaultmidpunct}
{\mcitedefaultendpunct}{\mcitedefaultseppunct}\relax
\EndOfBibitem
\bibitem[Smith \latin{et~al.}(2020)Smith, Burns, Simmonett, Parrish, Schieber, Galvelis, Kraus, Kruse, Di~Remigio, Alenaizan, James, Lehtola, Misiewicz, Scheurer, Shaw, Schriber, Xie, Glick, Sirianni, O’Brien, Waldrop, Kumar, Hohenstein, Pritchard, Brooks, Schaefer, Sokolov, Patkowski, DePrince, Bozkaya, King, Evangelista, Turney, Crawford, and Sherrill]{Psi4}
Smith,~D. G.~A.; Burns,~L.~A.; Simmonett,~A.~C.; Parrish,~R.~M.; Schieber,~M.~C.; Galvelis,~R.; Kraus,~P.; Kruse,~H.; Di~Remigio,~R.; Alenaizan,~A.; James,~A.~M.; Lehtola,~S.; Misiewicz,~J.~P.; Scheurer,~M.; Shaw,~R.~A.; Schriber,~J.~B.; Xie,~Y.; Glick,~Z.~L.; Sirianni,~D.~A.; O’Brien,~J.~S.; Waldrop,~J.~M.; Kumar,~A.; Hohenstein,~E.~G.; Pritchard,~B.~P.; Brooks,~B.~R.; Schaefer,~I.,~Henry~F.; Sokolov,~A.~Y.; Patkowski,~K.; DePrince,~I.,~A.~Eugene; Bozkaya,~U.; King,~R.~A.; Evangelista,~F.~A.; Turney,~J.~M.; Crawford,~T.~D.; Sherrill,~C.~D. {PSI4 1.4: Open-source software for high-throughput quantum chemistry}. \emph{The Journal of Chemical Physics} \textbf{2020}, \emph{152}, 184108\relax
\mciteBstWouldAddEndPuncttrue
\mciteSetBstMidEndSepPunct{\mcitedefaultmidpunct}
{\mcitedefaultendpunct}{\mcitedefaultseppunct}\relax
\EndOfBibitem
\bibitem[Folkestad \latin{et~al.}(2020)Folkestad, Kj{\o}nstad, Myhre, Andersen, Balbi, Coriani, Giovannini, Goletto, Haugland, Hutcheson, H{\o}yvik, Moitra, Paul, Scavino, Skeidsvoll, Tveten, and Koch]{eTprog}
Folkestad,~S.~D.; Kj{\o}nstad,~E.~F.; Myhre,~R.~H.; Andersen,~J.~H.; Balbi,~A.; Coriani,~S.; Giovannini,~T.; Goletto,~L.; Haugland,~T.~S.; Hutcheson,~A.; H{\o}yvik,~I.-M.; Moitra,~T.; Paul,~A.~C.; Scavino,~M.; Skeidsvoll,~A.~S.; Tveten,~{\AA}.~H.; Koch,~H. {e$^T$ 1.0: An open source electronic structure program with emphasis on coupled cluster and multilevel methods}. \emph{The Journal of Chemical Physic} \textbf{2020}, \emph{152}, 184103\relax
\mciteBstWouldAddEndPuncttrue
\mciteSetBstMidEndSepPunct{\mcitedefaultmidpunct}
{\mcitedefaultendpunct}{\mcitedefaultseppunct}\relax
\EndOfBibitem
\bibitem[Rinkevicius \latin{et~al.}(2020)Rinkevicius, Li, Vahtras, Ahmadzadeh, Brand, Ringholm, List, Scheurer, Scott, Dreuw, and Norman]{veloxchem}
Rinkevicius,~Z.; Li,~X.; Vahtras,~O.; Ahmadzadeh,~K.; Brand,~M.; Ringholm,~M.; List,~N.~H.; Scheurer,~M.; Scott,~M.; Dreuw,~A.; Norman,~P. VeloxChem: A Python-driven density-functional theory program for spectroscopy simulations in high-performance computing environments. \emph{WIREs Computational Molecular Science} \textbf{2020}, \emph{10}, e1457\relax
\mciteBstWouldAddEndPuncttrue
\mciteSetBstMidEndSepPunct{\mcitedefaultmidpunct}
{\mcitedefaultendpunct}{\mcitedefaultseppunct}\relax
\EndOfBibitem
\bibitem[Sa{\l}ek \latin{et~al.}(2002)Sa{\l}ek, Vahtras, Helgaker, and {\AA}gren]{Salek2002}
Sa{\l}ek,~P.; Vahtras,~O.; Helgaker,~T.; {\AA}gren,~H. {Density-functional theory of linear and nonlinear time-dependent molecular properties}. \emph{The Journal of Chemical Physics} \textbf{2002}, \emph{117}, 9630--9645\relax
\mciteBstWouldAddEndPuncttrue
\mciteSetBstMidEndSepPunct{\mcitedefaultmidpunct}
{\mcitedefaultendpunct}{\mcitedefaultseppunct}\relax
\EndOfBibitem
\bibitem[Norman(2011)]{Norman2011perspective}
Norman,~P. A perspective on nonresonant and resonant electronic response theory for time-dependent molecular properties. \emph{Physical Chemistry Chemical Physics} \textbf{2011}, \emph{13}, 20519\relax
\mciteBstWouldAddEndPuncttrue
\mciteSetBstMidEndSepPunct{\mcitedefaultmidpunct}
{\mcitedefaultendpunct}{\mcitedefaultseppunct}\relax
\EndOfBibitem
\bibitem[Zelevinsky \latin{et~al.}(2023)Zelevinsky, Izmaylov, and Alexandrova]{Zelevinsky2023}
Zelevinsky,~T.; Izmaylov,~A.~F.; Alexandrova,~A.~N. Physical Chemistry of Quantum Information Science. \emph{The Journal of Physical Chemistry A} \textbf{2023}, \emph{127}, 10357--10359\relax
\mciteBstWouldAddEndPuncttrue
\mciteSetBstMidEndSepPunct{\mcitedefaultmidpunct}
{\mcitedefaultendpunct}{\mcitedefaultseppunct}\relax
\EndOfBibitem
\bibitem[Aspuru-Guzik \latin{et~al.}(2005)Aspuru-Guzik, Dutoi, Love, and Head-Gordon]{Aspuru-Guzik2005}
Aspuru-Guzik,~A.; Dutoi,~A.~D.; Love,~P.~J.; Head-Gordon,~M. {Simulated Quantum Computation of Molecular Energies}. \emph{Science} \textbf{2005}, \emph{309}, 1704--1707\relax
\mciteBstWouldAddEndPuncttrue
\mciteSetBstMidEndSepPunct{\mcitedefaultmidpunct}
{\mcitedefaultendpunct}{\mcitedefaultseppunct}\relax
\EndOfBibitem
\bibitem[Cao \latin{et~al.}(2019)Cao, Romero, Olson, Degroote, Johnson, Kieferová, Kivlichan, Menke, Peropadre, Sawaya, Sim, Veis, and Aspuru-Guzik]{Cao2019}
Cao,~Y.; Romero,~J.; Olson,~J.~P.; Degroote,~M.; Johnson,~P.~D.; Kieferová,~M.; Kivlichan,~I.~D.; Menke,~T.; Peropadre,~B.; Sawaya,~N. P.~D.; Sim,~S.; Veis,~L.; Aspuru-Guzik,~A. Quantum Chemistry in the Age of Quantum Computing. \emph{Chemical Reviews} \textbf{2019}, \emph{119}, 10856--10915\relax
\mciteBstWouldAddEndPuncttrue
\mciteSetBstMidEndSepPunct{\mcitedefaultmidpunct}
{\mcitedefaultendpunct}{\mcitedefaultseppunct}\relax
\EndOfBibitem
\bibitem[Bauer \latin{et~al.}(2020)Bauer, Bravyi, Motta, and Chan]{Bauer2020}
Bauer,~B.; Bravyi,~S.; Motta,~M.; Chan,~G. K.-L. Quantum Algorithms for Quantum Chemistry and Quantum Materials Science. \emph{Chemical Reviews} \textbf{2020}, \emph{120}, 12685--12717\relax
\mciteBstWouldAddEndPuncttrue
\mciteSetBstMidEndSepPunct{\mcitedefaultmidpunct}
{\mcitedefaultendpunct}{\mcitedefaultseppunct}\relax
\EndOfBibitem
\bibitem[Chen \latin{et~al.}(2023)Chen, Cotler, Huang, and Li]{chen_complexity_2023}
Chen,~S.; Cotler,~J.; Huang,~H.-Y.; Li,~J. The complexity of {NISQ}. \emph{Nature Communications} \textbf{2023}, \emph{14}, 6001\relax
\mciteBstWouldAddEndPuncttrue
\mciteSetBstMidEndSepPunct{\mcitedefaultmidpunct}
{\mcitedefaultendpunct}{\mcitedefaultseppunct}\relax
\EndOfBibitem
\bibitem[McClean \latin{et~al.}(2016)McClean, Romero, Babbush, and Aspuru-Guzik]{McClean2016}
McClean,~J.~R.; Romero,~J.; Babbush,~R.; Aspuru-Guzik,~A. {The theory of variational hybrid quantum-classical algorithms}. \emph{New Journal of Physics} \textbf{2016}, \emph{18}, 023023\relax
\mciteBstWouldAddEndPuncttrue
\mciteSetBstMidEndSepPunct{\mcitedefaultmidpunct}
{\mcitedefaultendpunct}{\mcitedefaultseppunct}\relax
\EndOfBibitem
\bibitem[Ollitrault \latin{et~al.}(2020)Ollitrault, Kandala, Chen, Barkoutsos, Mezzacapo, Pistoia, Sheldon, Woerner, Gambetta, and Tavernelli]{Ollitrault2020}
Ollitrault,~P.~J.; Kandala,~A.; Chen,~C.-F.; Barkoutsos,~P.~K.; Mezzacapo,~A.; Pistoia,~M.; Sheldon,~S.; Woerner,~S.; Gambetta,~J.~M.; Tavernelli,~I. Quantum equation of motion for computing molecular excitation energies on a noisy quantum processor. \emph{Physical Review Research} \textbf{2020}, \emph{2}\relax
\mciteBstWouldAddEndPuncttrue
\mciteSetBstMidEndSepPunct{\mcitedefaultmidpunct}
{\mcitedefaultendpunct}{\mcitedefaultseppunct}\relax
\EndOfBibitem
\bibitem[Asthana \latin{et~al.}(2023)Asthana, Kumar, Abraham, Grimsley, Zhang, Cincio, Tretiak, Dub, Economou, Barnes, and Mayhall]{Asthana2023}
Asthana,~A.; Kumar,~A.; Abraham,~V.; Grimsley,~H.; Zhang,~Y.; Cincio,~L.; Tretiak,~S.; Dub,~P.~A.; Economou,~S.~E.; Barnes,~E.; Mayhall,~N.~J. Quantum self-consistent equation-of-motion method for computing molecular excitation energies, ionization potentials, and electron affinities on a quantum computer. \emph{Chemical Science} \textbf{2023}, \emph{14}, 2405--2418\relax
\mciteBstWouldAddEndPuncttrue
\mciteSetBstMidEndSepPunct{\mcitedefaultmidpunct}
{\mcitedefaultendpunct}{\mcitedefaultseppunct}\relax
\EndOfBibitem
\bibitem[Kumar \latin{et~al.}(2023)Kumar, Asthana, Abraham, Crawford, Mayhall, Zhang, Cincio, Tretiak, and Dub]{Kumar2023}
Kumar,~A.; Asthana,~A.; Abraham,~V.; Crawford,~T.~D.; Mayhall,~N.~J.; Zhang,~Y.; Cincio,~L.; Tretiak,~S.; Dub,~P.~A. Quantum Simulation of Molecular Response Properties in the NISQ Era. \emph{Journal of Chemical Theory and Computation} \textbf{2023}, \relax
\mciteBstWouldAddEndPunctfalse
\mciteSetBstMidEndSepPunct{\mcitedefaultmidpunct}
{}{\mcitedefaultseppunct}\relax
\EndOfBibitem
\bibitem[Kim and Krylov(2023)Kim, and Krylov]{Kim2023}
Kim,~Y.; Krylov,~A.~I. Two Algorithms for Excited-State Quantum Solvers: Theory and Application to EOM-UCCSD. \emph{The Journal of Physical Chemistry A} \textbf{2023}, \emph{127}, 6552--6566\relax
\mciteBstWouldAddEndPuncttrue
\mciteSetBstMidEndSepPunct{\mcitedefaultmidpunct}
{\mcitedefaultendpunct}{\mcitedefaultseppunct}\relax
\EndOfBibitem
\bibitem[Castellanos \latin{et~al.}(2023)Castellanos, Motta, and Rice]{Castellanos2023}
Castellanos,~M.~A.; Motta,~M.; Rice,~J.~E. Quantum computation of $\pi \to \pi^\ast$ and $\mathrm{n} \to \pi^\ast$ excited states of aromatic heterocycles. \emph{Molecular Physics} \textbf{2023}, \emph{0}, e2282736\relax
\mciteBstWouldAddEndPuncttrue
\mciteSetBstMidEndSepPunct{\mcitedefaultmidpunct}
{\mcitedefaultendpunct}{\mcitedefaultseppunct}\relax
\EndOfBibitem
\bibitem[Nakagawa \latin{et~al.}(2023)Nakagawa, Chen, Sudo, Ohnishi, and Mizukami]{nakagawa_analytical_2023}
Nakagawa,~Y.~O.; Chen,~J.; Sudo,~S.; Ohnishi,~Y.-y.; Mizukami,~W. Analytical Formulation of the Second-Order Derivative of Energy for the Orbital-Optimized Variational Quantum Eigensolver: Application to Polarizability. \emph{Journal of Chemical Theory and Computation} \textbf{2023}, \emph{19}, 1998--2009\relax
\mciteBstWouldAddEndPuncttrue
\mciteSetBstMidEndSepPunct{\mcitedefaultmidpunct}
{\mcitedefaultendpunct}{\mcitedefaultseppunct}\relax
\EndOfBibitem
\bibitem[Jensen \latin{et~al.}(2023)Jensen, Kjellgren, Reinholdt, Ziems, Coriani, Kongsted, and Sauer]{Jensen_qEOM_2023}
Jensen,~P. W.~K.; Kjellgren,~E.~R.; Reinholdt,~P.; Ziems,~K.~M.; Coriani,~S.; Kongsted,~J.; Sauer,~S. P.~A. Quantum Equation of Motion with Orbital Optimization for Computing Molecular Properties in Near-Term Quantum Computing. \emph{arXiv:2312.12386} \textbf{2023}, \relax
\mciteBstWouldAddEndPunctfalse
\mciteSetBstMidEndSepPunct{\mcitedefaultmidpunct}
{}{\mcitedefaultseppunct}\relax
\EndOfBibitem
\bibitem[Mizukami \latin{et~al.}(2020)Mizukami, Mitarai, Nakagawa, Yamamoto, Yan, and Ohnishi]{Mizukami2020}
Mizukami,~W.; Mitarai,~K.; Nakagawa,~Y.~O.; Yamamoto,~T.; Yan,~T.; Ohnishi,~Y. Orbital optimized unitary coupled cluster theory for quantum computer. \emph{Physical Review Research} \textbf{2020}, \emph{2}, 033421\relax
\mciteBstWouldAddEndPuncttrue
\mciteSetBstMidEndSepPunct{\mcitedefaultmidpunct}
{\mcitedefaultendpunct}{\mcitedefaultseppunct}\relax
\EndOfBibitem
\bibitem[Sokolov \latin{et~al.}(2020)Sokolov, Barkoutsos, Ollitrault, Greenberg, Rice, Pistoia, and Tavernelli]{Sokolov2020}
Sokolov,~I.~O.; Barkoutsos,~P.~K.; Ollitrault,~P.~J.; Greenberg,~D.; Rice,~J.; Pistoia,~M.; Tavernelli,~I. Quantum orbital-optimized unitary coupled cluster methods in the strongly correlated regime: Can quantum algorithms outperform their classical equivalents? \emph{The Journal of Chemical Physics} \textbf{2020}, \emph{152}\relax
\mciteBstWouldAddEndPuncttrue
\mciteSetBstMidEndSepPunct{\mcitedefaultmidpunct}
{\mcitedefaultendpunct}{\mcitedefaultseppunct}\relax
\EndOfBibitem
\bibitem[Fitzpatrick \latin{et~al.}(2022)Fitzpatrick, Nyk\"{a}nen, Talarico, Lunghi, Maniscalco, García-Pérez, and Knecht]{Fitzpatrick2022}
Fitzpatrick,~A.; Nyk\"{a}nen,~A.; Talarico,~N.~W.; Lunghi,~A.; Maniscalco,~S.; García-Pérez,~G.; Knecht,~S. A self-consistent field approach for the variational quantum eigensolver: orbital optimization goes adaptive. \textbf{2022}, \relax
\mciteBstWouldAddEndPunctfalse
\mciteSetBstMidEndSepPunct{\mcitedefaultmidpunct}
{}{\mcitedefaultseppunct}\relax
\EndOfBibitem
\bibitem[Helgaker \latin{et~al.}(2013)Helgaker, J{\o}rgensen, and Olsen]{Helgaker2013book}
Helgaker,~T.; J{\o}rgensen,~P.; Olsen,~J. \emph{Molecular electronic-structure theory}; John Wiley \& Sons: Nashville, TN, 2013\relax
\mciteBstWouldAddEndPuncttrue
\mciteSetBstMidEndSepPunct{\mcitedefaultmidpunct}
{\mcitedefaultendpunct}{\mcitedefaultseppunct}\relax
\EndOfBibitem
\bibitem[Siegbahn \latin{et~al.}(1980)Siegbahn, Heiberg, Roos, and Levy]{Siegbahn1980}
Siegbahn,~P.; Heiberg,~A.; Roos,~B.; Levy,~B. A Comparison of the Super-{CI} and the Newton-Raphson Scheme in the Complete Active Space {SCF} Method. \emph{Physica Scripta} \textbf{1980}, \emph{21}, 323--327\relax
\mciteBstWouldAddEndPuncttrue
\mciteSetBstMidEndSepPunct{\mcitedefaultmidpunct}
{\mcitedefaultendpunct}{\mcitedefaultseppunct}\relax
\EndOfBibitem
\bibitem[Roos \latin{et~al.}(1980)Roos, Taylor, and Sigbahn]{Roos1980}
Roos,~B.~O.; Taylor,~P.~R.; Sigbahn,~P.~E. A complete active space {SCF} method ({CASSCF}) using a density matrix formulated super-{CI} approach. \emph{Chemical Physics} \textbf{1980}, \emph{48}, 157--173\relax
\mciteBstWouldAddEndPuncttrue
\mciteSetBstMidEndSepPunct{\mcitedefaultmidpunct}
{\mcitedefaultendpunct}{\mcitedefaultseppunct}\relax
\EndOfBibitem
\bibitem[Siegbahn \latin{et~al.}(1981)Siegbahn, Alml\"{o}f, Heiberg, and Roos]{Siegbahn1981}
Siegbahn,~P. E.~M.; Alml\"{o}f,~J.; Heiberg,~A.; Roos,~B.~O. The complete active space {SCF} ({CASSCF}) method in a Newton{\textendash}Raphson formulation with application to the {HNO} molecule. \emph{The Journal of Chemical Physics} \textbf{1981}, \emph{74}, 2384--2396\relax
\mciteBstWouldAddEndPuncttrue
\mciteSetBstMidEndSepPunct{\mcitedefaultmidpunct}
{\mcitedefaultendpunct}{\mcitedefaultseppunct}\relax
\EndOfBibitem
\bibitem[Grimsley \latin{et~al.}(2019)Grimsley, Economou, Barnes, and Mayhall]{Grimsley2019}
Grimsley,~H.~R.; Economou,~S.~E.; Barnes,~E.; Mayhall,~N.~J. An adaptive variational algorithm for exact molecular simulations on a quantum computer. \emph{Nature Communications} \textbf{2019}, \emph{10}\relax
\mciteBstWouldAddEndPuncttrue
\mciteSetBstMidEndSepPunct{\mcitedefaultmidpunct}
{\mcitedefaultendpunct}{\mcitedefaultseppunct}\relax
\EndOfBibitem
\bibitem[Tang \latin{et~al.}(2021)Tang, Shkolnikov, Barron, Grimsley, Mayhall, Barnes, and Economou]{Tang2021}
Tang,~H.~L.; Shkolnikov,~V.; Barron,~G.~S.; Grimsley,~H.~R.; Mayhall,~N.~J.; Barnes,~E.; Economou,~S.~E. Qubit-{ADAPT}-{VQE}: An Adaptive Algorithm for Constructing Hardware-Efficient Ans\"{a}tze on a Quantum Processor. \emph{{PRX} Quantum} \textbf{2021}, \emph{2}\relax
\mciteBstWouldAddEndPuncttrue
\mciteSetBstMidEndSepPunct{\mcitedefaultmidpunct}
{\mcitedefaultendpunct}{\mcitedefaultseppunct}\relax
\EndOfBibitem
\bibitem[J{\o}rgensen \latin{et~al.}(1988)J{\o}rgensen, Jensen, and Olsen]{Jorgensen1988}
J{\o}rgensen,~P.; Jensen,~H. J.~A.; Olsen,~J. Linear response calculations for large scale multiconfiguration self-consistent field wave functions. \emph{The Journal of Chemical Physics} \textbf{1988}, \emph{89}, 3654–3661\relax
\mciteBstWouldAddEndPuncttrue
\mciteSetBstMidEndSepPunct{\mcitedefaultmidpunct}
{\mcitedefaultendpunct}{\mcitedefaultseppunct}\relax
\EndOfBibitem
\bibitem[Paldus \latin{et~al.}(1977)Paldus, Adams, and Čížek]{Paldus1977}
Paldus,~J.; Adams,~B.~G.; Čížek,~J. Application of graphical methods of spin algebras to limited <scp>CI</scp> approaches. I. Closed shell case. \emph{International Journal of Quantum Chemistry} \textbf{1977}, \emph{11}, 813–848\relax
\mciteBstWouldAddEndPuncttrue
\mciteSetBstMidEndSepPunct{\mcitedefaultmidpunct}
{\mcitedefaultendpunct}{\mcitedefaultseppunct}\relax
\EndOfBibitem
\bibitem[Piecuch and Paldus(1989)Piecuch, and Paldus]{Piecuch1989}
Piecuch,~P.; Paldus,~J. Orthogonally spin‐adapted coupled‐cluster equations involving singly and doubly excited clusters. Comparison of different procedures for spin‐adaptation. \emph{International Journal of Quantum Chemistry} \textbf{1989}, \emph{36}, 429–453\relax
\mciteBstWouldAddEndPuncttrue
\mciteSetBstMidEndSepPunct{\mcitedefaultmidpunct}
{\mcitedefaultendpunct}{\mcitedefaultseppunct}\relax
\EndOfBibitem
\bibitem[Packer \latin{et~al.}(1996)Packer, Dalskov, Enevoldsen, Jensen, and Oddershede]{Packer1996}
Packer,~M.~J.; Dalskov,~E.~K.; Enevoldsen,~T.; Jensen,~H. J.~A.; Oddershede,~J. A new implementation of the second-order polarization propagator approximation ({SOPPA}): The excitation spectra of benzene and naphthalene. \emph{The Journal of Chemical Physics} \textbf{1996}, \emph{105}, 5886--5900\relax
\mciteBstWouldAddEndPuncttrue
\mciteSetBstMidEndSepPunct{\mcitedefaultmidpunct}
{\mcitedefaultendpunct}{\mcitedefaultseppunct}\relax
\EndOfBibitem
\bibitem[Rizzo \latin{et~al.}(2022)Rizzo, Libbi, Tacchino, Ollitrault, Marzari, and Tavernelli]{Rizzo2022}
Rizzo,~J.; Libbi,~F.; Tacchino,~F.; Ollitrault,~P.~J.; Marzari,~N.; Tavernelli,~I. One-particle Green{\textquotesingle}s functions from the quantum equation~of motion algorithm. \emph{Physical Review Research} \textbf{2022}, \emph{4}\relax
\mciteBstWouldAddEndPuncttrue
\mciteSetBstMidEndSepPunct{\mcitedefaultmidpunct}
{\mcitedefaultendpunct}{\mcitedefaultseppunct}\relax
\EndOfBibitem
\bibitem[Aharonov \latin{et~al.}(2006)Aharonov, Jones, and Landau]{aharonov2006polynomial}
Aharonov,~D.; Jones,~V.; Landau,~Z. A polynomial quantum algorithm for approximating the Jones polynomial. Proceedings of the Thirty-Eighth Annual ACM Symposium on Theory of Computing. 2006; pp 427--436\relax
\mciteBstWouldAddEndPuncttrue
\mciteSetBstMidEndSepPunct{\mcitedefaultmidpunct}
{\mcitedefaultendpunct}{\mcitedefaultseppunct}\relax
\EndOfBibitem
\bibitem[Knill \latin{et~al.}(2007)Knill, Ortiz, and Somma]{knill2007optimal}
Knill,~E.; Ortiz,~G.; Somma,~R.~D. Optimal quantum measurements of expectation values of observables. \emph{Physical Review A} \textbf{2007}, \emph{75}, 012328\relax
\mciteBstWouldAddEndPuncttrue
\mciteSetBstMidEndSepPunct{\mcitedefaultmidpunct}
{\mcitedefaultendpunct}{\mcitedefaultseppunct}\relax
\EndOfBibitem
\bibitem[Dob{\v{s}}{\'\i}{\v{c}}ek \latin{et~al.}(2007)Dob{\v{s}}{\'\i}{\v{c}}ek, Johansson, Shumeiko, and Wendin]{dobvsivcek2007arbitrary}
Dob{\v{s}}{\'\i}{\v{c}}ek,~M.; Johansson,~G.; Shumeiko,~V.; Wendin,~G. Arbitrary accuracy iterative quantum phase estimation algorithm using a single ancillary qubit: A two-qubit benchmark. \emph{Physical Review A} \textbf{2007}, \emph{76}, 030306\relax
\mciteBstWouldAddEndPuncttrue
\mciteSetBstMidEndSepPunct{\mcitedefaultmidpunct}
{\mcitedefaultendpunct}{\mcitedefaultseppunct}\relax
\EndOfBibitem
\bibitem[Parrish \latin{et~al.}(2019)Parrish, Hohenstein, McMahon, and Mart{\'{\i}}nez]{Parrish2019}
Parrish,~R.~M.; Hohenstein,~E.~G.; McMahon,~P.~L.; Mart{\'{\i}}nez,~T.~J. Quantum Computation of Electronic Transitions Using a Variational Quantum Eigensolver. \emph{Physical Review Letters} \textbf{2019}, \emph{122}\relax
\mciteBstWouldAddEndPuncttrue
\mciteSetBstMidEndSepPunct{\mcitedefaultmidpunct}
{\mcitedefaultendpunct}{\mcitedefaultseppunct}\relax
\EndOfBibitem
\bibitem[Nakanishi \latin{et~al.}(2019)Nakanishi, Mitarai, and Fujii]{Nakanishi2019}
Nakanishi,~K.~M.; Mitarai,~K.; Fujii,~K. Subspace-search variational quantum eigensolver for excited states. \emph{Physical Review Research} \textbf{2019}, \emph{1}\relax
\mciteBstWouldAddEndPuncttrue
\mciteSetBstMidEndSepPunct{\mcitedefaultmidpunct}
{\mcitedefaultendpunct}{\mcitedefaultseppunct}\relax
\EndOfBibitem
\bibitem[Pritchard \latin{et~al.}(2019)Pritchard, Altarawy, Didier, Gibson, and Windus]{Pritchard2019}
Pritchard,~B.~P.; Altarawy,~D.; Didier,~B.; Gibson,~T.~D.; Windus,~T.~L. New Basis Set Exchange: An Open, Up-to-Date Resource for the Molecular Sciences Community. \emph{Journal of Chemical Information and Modeling} \textbf{2019}, \emph{59}, 4814--4820\relax
\mciteBstWouldAddEndPuncttrue
\mciteSetBstMidEndSepPunct{\mcitedefaultmidpunct}
{\mcitedefaultendpunct}{\mcitedefaultseppunct}\relax
\EndOfBibitem
\bibitem[Feller(1996)]{Feller1996}
Feller,~D. The role of databases in support of computational chemistry calculations. \emph{Journal of Computational Chemistry} \textbf{1996}, \emph{17}, 1571--1586\relax
\mciteBstWouldAddEndPuncttrue
\mciteSetBstMidEndSepPunct{\mcitedefaultmidpunct}
{\mcitedefaultendpunct}{\mcitedefaultseppunct}\relax
\EndOfBibitem
\bibitem[Schuchardt \latin{et~al.}(2007)Schuchardt, Didier, Elsethagen, Sun, Gurumoorthi, Chase, Li, and Windus]{Schuchardt2007}
Schuchardt,~K.~L.; Didier,~B.~T.; Elsethagen,~T.; Sun,~L.; Gurumoorthi,~V.; Chase,~J.; Li,~J.; Windus,~T.~L. Basis Set Exchange:{\hspace{0.167em}} A Community Database for Computational Sciences. \emph{Journal of Chemical Information and Modeling} \textbf{2007}, \emph{47}, 1045--1052\relax
\mciteBstWouldAddEndPuncttrue
\mciteSetBstMidEndSepPunct{\mcitedefaultmidpunct}
{\mcitedefaultendpunct}{\mcitedefaultseppunct}\relax
\EndOfBibitem
\bibitem[Dill and Pople(1975)Dill, and Pople]{Dill1975}
Dill,~J.~D.; Pople,~J.~A. Self-consistent molecular orbital methods. {XV}. Extended Gaussian-type basis sets for lithium, beryllium, and boron. \emph{The Journal of Chemical Physics} \textbf{1975}, \emph{62}, 2921--2923\relax
\mciteBstWouldAddEndPuncttrue
\mciteSetBstMidEndSepPunct{\mcitedefaultmidpunct}
{\mcitedefaultendpunct}{\mcitedefaultseppunct}\relax
\EndOfBibitem
\bibitem[Ditchfield \latin{et~al.}(1971)Ditchfield, Hehre, and Pople]{Ditchfield1971}
Ditchfield,~R.; Hehre,~W.~J.; Pople,~J.~A. Self-Consistent Molecular-Orbital Methods. {IX}. An Extended Gaussian-Type Basis for Molecular-Orbital Studies of Organic Molecules. \emph{The Journal of Chemical Physics} \textbf{1971}, \emph{54}, 724--728\relax
\mciteBstWouldAddEndPuncttrue
\mciteSetBstMidEndSepPunct{\mcitedefaultmidpunct}
{\mcitedefaultendpunct}{\mcitedefaultseppunct}\relax
\EndOfBibitem
\bibitem[Hehre \latin{et~al.}(1972)Hehre, Ditchfield, and Pople]{Hehre1972}
Hehre,~W.~J.; Ditchfield,~R.; Pople,~J.~A. Self{\textemdash}Consistent Molecular Orbital Methods. {XII}. Further Extensions of Gaussian{\textemdash}Type Basis Sets for Use in Molecular Orbital Studies of Organic Molecules. \emph{The Journal of Chemical Physics} \textbf{1972}, \emph{56}, 2257--2261\relax
\mciteBstWouldAddEndPuncttrue
\mciteSetBstMidEndSepPunct{\mcitedefaultmidpunct}
{\mcitedefaultendpunct}{\mcitedefaultseppunct}\relax
\EndOfBibitem
\bibitem[Dunning(1989)]{Dunning1989}
Dunning,~T.~H. Gaussian basis sets for use in correlated molecular calculations. I. The atoms boron through neon and hydrogen. \emph{The Journal of Chemical Physics} \textbf{1989}, \emph{90}, 1007--1023\relax
\mciteBstWouldAddEndPuncttrue
\mciteSetBstMidEndSepPunct{\mcitedefaultmidpunct}
{\mcitedefaultendpunct}{\mcitedefaultseppunct}\relax
\EndOfBibitem
\bibitem[Prascher \latin{et~al.}(2010)Prascher, Woon, Peterson, Dunning, and Wilson]{Prascher2010}
Prascher,~B.~P.; Woon,~D.~E.; Peterson,~K.~A.; Dunning,~T.~H.; Wilson,~A.~K. Gaussian basis sets for use in correlated molecular calculations. {VII}. Valence, core-valence, and scalar relativistic basis sets for Li, Be, Na, and Mg. \emph{Theoretical Chemistry Accounts} \textbf{2010}, \emph{128}, 69--82\relax
\mciteBstWouldAddEndPuncttrue
\mciteSetBstMidEndSepPunct{\mcitedefaultmidpunct}
{\mcitedefaultendpunct}{\mcitedefaultseppunct}\relax
\EndOfBibitem
\bibitem[Kjellgren and Ziems()Kjellgren, and Ziems]{slowquant}
Kjellgren,~E.; Ziems,~K.~M. Slow{Q}uant. \url{https://github.com/erikkjellgren/SlowQuant/tree/master}\relax
\mciteBstWouldAddEndPuncttrue
\mciteSetBstMidEndSepPunct{\mcitedefaultmidpunct}
{\mcitedefaultendpunct}{\mcitedefaultseppunct}\relax
\EndOfBibitem
\bibitem[Sun(2015)]{Sun2015}
Sun,~Q. Libcint: An efficient general integral library for Gaussian basis functions. \emph{Journal of Computational Chemistry} \textbf{2015}, \emph{36}, 1664–1671\relax
\mciteBstWouldAddEndPuncttrue
\mciteSetBstMidEndSepPunct{\mcitedefaultmidpunct}
{\mcitedefaultendpunct}{\mcitedefaultseppunct}\relax
\EndOfBibitem
\bibitem[Sun \latin{et~al.}(2020)Sun, Zhang, Banerjee, Bao, Barbry, Blunt, Bogdanov, Booth, Chen, Cui, Eriksen, Gao, Guo, Hermann, Hermes, Koh, Koval, Lehtola, Li, Liu, Mardirossian, McClain, Motta, Mussard, Pham, Pulkin, Purwanto, Robinson, Ronca, Sayfutyarova, Scheurer, Schurkus, Smith, Sun, Sun, Upadhyay, Wagner, Wang, White, Whitfield, Williamson, Wouters, Yang, Yu, Zhu, Berkelbach, Sharma, Sokolov, and Chan]{Sun2020}
Sun,~Q.; Zhang,~X.; Banerjee,~S.; Bao,~P.; Barbry,~M.; Blunt,~N.~S.; Bogdanov,~N.~A.; Booth,~G.~H.; Chen,~J.; Cui,~Z.-H.; Eriksen,~J.~J.; Gao,~Y.; Guo,~S.; Hermann,~J.; Hermes,~M.~R.; Koh,~K.; Koval,~P.; Lehtola,~S.; Li,~Z.; Liu,~J.; Mardirossian,~N.; McClain,~J.~D.; Motta,~M.; Mussard,~B.; Pham,~H.~Q.; Pulkin,~A.; Purwanto,~W.; Robinson,~P.~J.; Ronca,~E.; Sayfutyarova,~E.~R.; Scheurer,~M.; Schurkus,~H.~F.; Smith,~J. E.~T.; Sun,~C.; Sun,~S.-N.; Upadhyay,~S.; Wagner,~L.~K.; Wang,~X.; White,~A.; Whitfield,~J.~D.; Williamson,~M.~J.; Wouters,~S.; Yang,~J.; Yu,~J.~M.; Zhu,~T.; Berkelbach,~T.~C.; Sharma,~S.; Sokolov,~A.~Y.; Chan,~G. K.-L. Recent developments in the PySCF program package. \emph{The Journal of Chemical Physics} \textbf{2020}, \emph{153}\relax
\mciteBstWouldAddEndPuncttrue
\mciteSetBstMidEndSepPunct{\mcitedefaultmidpunct}
{\mcitedefaultendpunct}{\mcitedefaultseppunct}\relax
\EndOfBibitem
\bibitem[M{\o}ller and Plesset(1934)M{\o}ller, and Plesset]{Mller1934}
M{\o}ller,~C.; Plesset,~M.~S. Note on an Approximation Treatment for Many-Electron Systems. \emph{Physical Review} \textbf{1934}, \emph{46}, 618--622\relax
\mciteBstWouldAddEndPuncttrue
\mciteSetBstMidEndSepPunct{\mcitedefaultmidpunct}
{\mcitedefaultendpunct}{\mcitedefaultseppunct}\relax
\EndOfBibitem
\bibitem[Jensen \latin{et~al.}(1988)Jensen, J{\o}rgensen, {\AA}gren, and Olsen]{Jensen1988}
Jensen,~H. J.~A.; J{\o}rgensen,~P.; {\AA}gren,~H.; Olsen,~J. Second-order M{\o}ller{\textendash}Plesset perturbation theory as a configuration and orbital generator in multiconfiguration self-consistent field calculations. \emph{The Journal of Chemical Physics} \textbf{1988}, \emph{88}, 3834--3839\relax
\mciteBstWouldAddEndPuncttrue
\mciteSetBstMidEndSepPunct{\mcitedefaultmidpunct}
{\mcitedefaultendpunct}{\mcitedefaultseppunct}\relax
\EndOfBibitem
\bibitem[Olsen \latin{et~al.}(2020)Olsen, Reine, Vahtras, Kjellgren, Reinholdt, Dundas, Li, Cukras, Ringholm, Hedeg{\aa}rd, Remigio, List, Faber, Tenorio, Bast, Pedersen, Rinkevicius, Sauer, Mikkelsen, Kongsted, Coriani, Ruud, Helgaker, Jensen, and Norman]{Olsen2020}
Olsen,~J. M.~H.; Reine,~S.; Vahtras,~O.; Kjellgren,~E.; Reinholdt,~P.; Dundas,~K. O.~H.; Li,~X.; Cukras,~J.; Ringholm,~M.; Hedeg{\aa}rd,~E.~D.; Remigio,~R.~D.; List,~N.~H.; Faber,~R.; Tenorio,~B. N.~C.; Bast,~R.; Pedersen,~T.~B.; Rinkevicius,~Z.; Sauer,~S. P.~A.; Mikkelsen,~K.~V.; Kongsted,~J.; Coriani,~S.; Ruud,~K.; Helgaker,~T.; Jensen,~H. J.~A.; Norman,~P. Dalton Project: A Python platform for molecular- and electronic-structure simulations of complex systems. \emph{The Journal of Chemical Physics} \textbf{2020}, \emph{152}, 214115\relax
\mciteBstWouldAddEndPuncttrue
\mciteSetBstMidEndSepPunct{\mcitedefaultmidpunct}
{\mcitedefaultendpunct}{\mcitedefaultseppunct}\relax
\EndOfBibitem
\bibitem[Wang and Song(2016)Wang, and Song]{wang2016geometry}
Wang,~L.-P.; Song,~C. Geometry optimization made simple with translation and rotation coordinates. \emph{The Journal of Chemical Physics} \textbf{2016}, \emph{144}, 214108\relax
\mciteBstWouldAddEndPuncttrue
\mciteSetBstMidEndSepPunct{\mcitedefaultmidpunct}
{\mcitedefaultendpunct}{\mcitedefaultseppunct}\relax
\EndOfBibitem
\bibitem[Levine \latin{et~al.}(2008)Levine, Coe, and Mart{\'\i}nez]{levine2008optimizing}
Levine,~B.~G.; Coe,~J.~D.; Mart{\'\i}nez,~T.~J. Optimizing conical intersections without derivative coupling vectors: Application to multistate multireference second-order perturbation theory (MS-CASPT2). \emph{The Journal of Physical Chemistry B} \textbf{2008}, \emph{112}, 405--413\relax
\mciteBstWouldAddEndPuncttrue
\mciteSetBstMidEndSepPunct{\mcitedefaultmidpunct}
{\mcitedefaultendpunct}{\mcitedefaultseppunct}\relax
\EndOfBibitem
\end{mcitethebibliography}

\end{document}